\documentclass[twocolumn]{aastex631}  
\usepackage{natbib}
\usepackage{amsmath}
\usepackage[figuresright]{rotating}
\usepackage{url}
\usepackage{longtable}
\usepackage{newunicodechar}
\usepackage{xcolor} 
\usepackage{booktabs}
\usepackage{booktabs}
\usepackage{xcolor}

\newunicodechar{−}{-}

\begin{document}
	
\title{Fast Optical Variability of the TeV Blazar PKS 1725+123 Observed by SVOM-VT and Insights from Multiwavelength Follow-up Observations}
	
\correspondingauthor{Alexis Coleiro and Jin Zhang}
\email{coleiro@apc.in2p3.fr, j.zhang@bit.edu.cn}

\author{Shuo-Yu Liu}
\affiliation{School of Physics, Beijing Institute of Technology, Beijing 100081, People's Republic of China; j.zhang@bit.edu.cn}

\author{Yu-Wei Yu}
\affiliation{School of Physics, Beijing Institute of Technology, Beijing 100081, People's Republic of China; j.zhang@bit.edu.cn}

\author{Ji-Shun Lian}
\affiliation{School of Physics, Beijing Institute of Technology, Beijing 100081, People's Republic of China; j.zhang@bit.edu.cn}

\author{Xin-Ke Hu}
\affiliation{School of Physics, Beijing Institute of Technology, Beijing 100081, People's Republic of China; j.zhang@bit.edu.cn}

\author{Alexis Coleiro\dag}
\affiliation{Université Paris Cite, CNRS, Astroparticule et Cosmologie, F-75013 Paris, France; coleiro@apc.in2p3.fr}

\author{Zhu-Heng Yao}
\affiliation{National Astronomical Observatories, Chinese Academy of Sciences, Beijing 100101, People’s Republic of China}

\author{Li-Ping Xin}
\affiliation{National Astronomical Observatories, Chinese Academy of Sciences, Beijing 100101, People’s Republic of China}

\author{Jing Wang}
\affiliation{National Astronomical Observatories, Chinese Academy of Sciences, Beijing 100101, People’s Republic of China}

\author{Hua-Li Li}
\affiliation{National Astronomical Observatories, Chinese Academy of Sciences, Beijing 100101, People’s Republic of China}

\author{Zi-Qi Wang} 
\affiliation{Guangxi Key Laboratory for Relativistic Astrophysics, School of Physical Science and Technology, Guangxi University, Nanning 530004, People’s Republic of China}

\author{Jin Zhang\dag}
\affiliation{School of Physics, Beijing Institute of Technology, Beijing 100081, People's Republic of China; j.zhang@bit.edu.cn}

\author{Floriane Cangemi}
\affiliation{Université Paris Cite, CNRS, Astroparticule et Cosmologie, F-75013 Paris, France; coleiro@apc.in2p3.fr}

\author{Bertrand Cordier}
\affiliation{CEA Paris-Saclay, Institut de Recherche sur les lois Fondamentales de l'Univers, 9111 Gif-sur-Yvette, France}

\author{Antoine Foisseau}
\affiliation{Aix Marseille Univ, CNRS/IN2P3, CPPM, Marseille, France}

\author{Olivier Godet}
\affiliation{IRAP, Université de Toulouse, CNRS, CNES, Toulouse, France}

\author{Andrea Goldwurm}
\affiliation{Université Paris Cite, CNRS, Astroparticule et Cosmologie, F-75013 Paris, France; coleiro@apc.in2p3.fr}\affiliation{CEA Paris-Saclay, Institut de Recherche sur les lois Fondamentales de l'Univers, 9111 Gif-sur-Yvette, France}

\author{Diego Götz}
\affiliation{CEA Paris-Saclay, Institut de Recherche sur les lois Fondamentales de l'Univers, 9111 Gif-sur-Yvette, France}

\author{Sébastien Guillot}
\affiliation{IRAP, Université de Toulouse, CNRS, CNES, Toulouse, France}

\author{Xu-Hui Han}
\affiliation{National Astronomical Observatories, Chinese Academy of Sciences, Beijing 100101, People’s Republic of China}

\author{Ning Jiang}
\affiliation{Department of Astronomy, University of Science and Technology of China, Hefei 230026, People’s Republic of China (2) School of Astronomy and Space Sciences, University of Science and Technology of China, Hefei 230026, People’s Republic of China}

\author{Cyril Lachaud}
\affiliation{Université Paris Cite, CNRS, Astroparticule et Cosmologie, F-75013 Paris, France; coleiro@apc.in2p3.fr}

\author{Sébastien Le Stum}
\affiliation{Universite Paris Cite, CNRS, Astroparticule et Cosmologie, F-75013 Paris, France}

\author{En-Wei Liang}
\affiliation{Guangxi Key Laboratory for Relativistic Astrophysics, School of Physical Science and Technology, Guangxi University, Nanning 530004, People’s Republic of China}

\author{Pierre Maggi}
\affiliation{Observatoire Astronomique de Strasbourg, Université de Strasbourg, CNRS, 11 rue de l’Université, F-67000
Strasbourg, France}

\author{Yu-Lei Qiu}
\affiliation{National Astronomical Observatories, Chinese Academy of Sciences, Beijing 100101, People’s Republic of China}

\author{Jérôme Rodriguez}
\affiliation{Université Paris-Saclay, Université Paris Cité, CEA, CNRS, AIM, 91191 Gif-sur-Yvette, France}
\affiliation{Observatoire des Sciences de l'Univers de l’Université Paris Saclay,  Université Paris-Saclay, CNRS, Bât. 121, 91405 Orsay, France}

\author{Lian Tao}
\affiliation{State Key Laboratory of Particle Astrophysics, Institute of High Energy Physics, Chinese Academy of Sciences, Beijing 100049, People’s Republic of China}

\author{Jian-Yan Wei}
\affiliation{National Astronomical Observatories, Chinese Academy of Sciences, Beijing 100101, People’s Republic of China}\affiliation{School of Astronomy and Space Science, University of Chinese Academy of Sciences, Beijing 100049, People’s Republic of China}

\author{Chao Wu}
\affiliation{National Astronomical Observatories, Chinese Academy of Sciences, Beijing 100101, People’s Republic of China}

\author{Liang Zhang}
\affiliation{State Key Laboratory of Particle Astrophysics, Institute of High Energy Physics, Chinese Academy of Sciences, Beijing 100049, People’s Republic of China}

\author{Shuang-Nan Zhang}
\affiliation{State Key Laboratory of Particle Astrophysics, Institute of High Energy Physics, Chinese Academy of Sciences, Beijing 100049, People’s Republic of China}

\author{Shi-Jie Zheng}
\affiliation{State Key Laboratory of Particle Astrophysics, Institute of High Energy Physics, Chinese Academy of Sciences, Beijing 100049, People’s Republic of China}

\begin{abstract}

PKS 1725+123 is a flat-spectrum radio quasar (FSRQ) with a redshift of $z=0.586$. The detection of this object in the TeV band was reported by the MAGIC telescopes and H.E.S.S. in August 2025. Subsequently, we promptly initiated Target-of-Opportunity observations using the Space-based multi-band astronomical Variable Objects Monitor (SVOM) satellite. By analyzing the observational optical data from SVOM-VT and comprehensively examining the Fermi-LAT and Swift-XRT observational data, it was found that the source is in a high-flux state across the optical, X-ray, and GeV $\gamma$-ray bands around the time of the TeV detections. Its optical flux reaches a historically unprecedented high level and shows significant variability on timescale as short as minutes. The variability is accompanied by changes in the color index, exhibiting a \textit{bluer when brighter} behavior during the high-flux state. Based on the simultaneous Multiwavelength data, we construct the broadband spectral energy distribution (SED) of the source in the high-flux state. PKS 1725+123 demonstrates a remarkably high synchrotron peak frequency, which is distinctly different from that of other FSRQs. We propose a two-zone spine-sheath jet model to reproduce this SED. The optical--X-ray emission is generated by the synchrotron process of the relativistic electrons within a compact zone. The inverse Compton (IC) scattering processes of the same electron population contribute to the low-energy end of the Fermi-LAT spectrum, while the high-energy end of the Fermi-LAT spectrum is ascribed to the IC scattering of the synchrotron photons within the compact zone by the higher-energy electrons in an extended region.   

\end{abstract}


\section{Introduction}

Blazars are a subclass of radio-loud active galactic nuclei (AGNs), in which their jets are nearly aligned with the line of sight \citep{1995PASP..107..803U}. Blazars are further divided into flat-spectrum radio quasars (FSRQs) and BL Lacertae objects according to the presence or absence of prominent optical emission lines \citep{1991ApJ...374..431S, 1991ApJS...76..813S}. The emission of blazars, which spans the entire electromagnetic spectrum from radio to $\gamma$-ray band, is dominated by their relativistic jets. The broadband spectral energy distributions (SEDs) of blazars usually exhibit a bimodal pattern. The low-energy hump is produced by the synchrotron radiation of relativistic electrons in the jets, while the high-energy hump is generally attributed to the inverse Compton (IC) scattering of relativistic electrons \citep{1992ApJ...397L...5M,1993ApJ...416..458D,1996MNRAS.280...67G,2009ApJ...704...38S,2012ApJ...752..157Z,2014ApJ...788..104Z, 2015ApJ...807...51Z}. 

Blazars are characterized by violent variability in multiple wavelengths. The timescales of flux variations for blazars range from minutes to years \citep{1997ARA&A..35..445U,2007Ap&SS.309...95B,2008ApJ...677..906F}. In the most extreme cases, the variability timescales at $\gamma$-ray band can be as short as a few minutes, such as in Mrk 421 \citep{1996Natur.383..319G}, Mrk 501 \citep{2007ApJ...669..862A}, PKS 2155--304 \citep{2007ApJ...664L..71A}, BL Lacertae \citep{2013ApJ...762...92A}, and FSRQ PKS 1222+21 \citep{2011ApJ...730L...8A}. Rapid and violent flux variations indicate a small and dense radiation region and particle acceleration or energy injection in extreme physical processes. The flux variations are commonly accompanied by the spectral evolutions \citep[e.g.,][]{2026ApJ..1001...97H}, and the different spectral variation features may indicate distinctly dominant particle acceleration mechanisms (e.g., \citealt{2008A&A...478..395M,2011ApJ...739...66T}). Blazars are the confirmed primary extragalactic $\gamma$-ray emitting objects in the GeV-TeV bands. More than 3900 blazars and blazar candidates have been reported as $\gamma$-ray emitting sources in the 14-year Fermi-LAT catalog (4FGL-DR4, \citealt{Ballet2024_4FGLDR4}). In particular, the majority of the detected TeV extragalactic objects are blazars, approximately 90 blazars in the TeVCat catalog\footnote{\url{http://tevcat2.uchicago.edu/}}. However, among these TeV blazars, fewer than 10 are FSRQs, while the remainder are BL Lac objects. 

PKS 1725+123 is a $\gamma$-ray emitting FSRQ located at a redshift of $z = 0.586$ \citep{2013ApJ...764..135S}. On 2025 August 19, it was reported that PKS 1725+123 was detected in the very high energy (VHE) TeV band by the Major Atmospheric Gamma-ray Imaging Cherenkov (MAGIC) telescopes \citep{2025ATel17344....1P}. Subsequently, it was also reported to have been detected by the High Energy Stereoscopic System (H.E.S.S.; \citealt{2025ATel17346....1W}). The Space-based multi-band astronomical Variable Objects Monito (SVOM; \citealt{2016arXiv161006892W,cordier2026svom_mission}) promptly conducted follow-up Target-of-Opportunity (ToO) observations. In this letter, Section \ref{sec:svom} describes the SVOM observations and data analysis for PKS 1725+123. Section \ref{sec:result} presents the SVOM observation results, together with the simultaneous observation results of Swift-XRT and Fermi-LAT. Based on the simultaneous Multiwavelength observation data, Section \ref{sec:SED} constructs and fits the broadband SED of PKS 1725+123. Section \ref{sec:dis&con} provides discussion and conclusions.

\section{SVOM Observations and Data Analysis} \label{sec:svom}

There are four detectors on board the SVOM, namely the Gamma-Ray Burst Monitor (GRM), the ECLAIRs hard X-ray coded-mask telescope, the Multi-channel X-ray telescope (MXT), and the Visible Telescope (VT). We carried out Target-of-Opportunity (ToO) observations of PKS 1725+123 using SVOM from 2025 August 20 to 2025 September 8, before the source became inaccessible due to solar constraints. ToO observations are conducted in the framework of the SVOM observatory science program, which dedicates part of the non-Gamma Ray Burst observation time of SVOM to the monitoring of blazars \citep{2026A&A...707A.349F,qiu2026svom_vt_overview,Coleiro2026}. The SVOM-VT detector features two bands: $R$-band (VT$\_R$, $\lambda$ = 6500~\AA, FWHM = 1300~\AA) and $B$-band (VT$\_B$, $\lambda$ = 4500~\AA, FWHM = 1000~\AA), where $\lambda$ = 6500~\AA\,and $\lambda$ = 4500~\AA\, are the central wavelengths. The MXT detector (\citealt{2023ExA....55..487G}) is a soft X-ray (0.3--10\,keV) "lobster-eye" telescope on board the SVOM, while the ECLAIRs (\citealt{godet2026eclairs}) covers the 4--150 keV energy range. 

The detailed observational orbital parameters are listed in Table \ref{tab:svom}. The target was monitored for approximately 45 min during each satellite orbit. The exposure time was set to 50 s for each individual image of SVOM-VT observations. We used the Level-2 data provided by the SVOM-VT science team, which had been corrected for overscan, bias, dark current, and flat-field effects \citep{han2026svom_suss,yao2026svom_vt_calibration}. All images were processed individually, and images with a median background level exceeding 300 photons per pixel were excluded. Aperture photometry was performed using \texttt{Photutils v2.3.0} \citep{larry_bradley_2025_14889440}, an \texttt{Astropy} package designed for the detection and photometric measurement of astronomical sources. Using an aperture-growth method, we selected the aperture radius that maximized the signal-to-noise ratio as the photometric aperture, yielding a radius of about 2.0 pixels. Given that the aperture radius employed for photometry is insufficient to encompass the total flux of sources, aperture correction was considered and derived using a 10-pixel-radius aperture. For each image, all isolated sources brighter than 18 mag were measured using both aperture radii, and the mean correction was determined via an iterative process. The aperture-corrected fluxes were converted into magnitudes by employing photometric zero points of 23.02 mag and 23.10 mag for the VT$\_R$ and VT$\_B$ bands, respectively. The photometric uncertainties and aperture-correction errors were propagated to estimate the final magnitudes and their uncertainties \citep{li2026svom_vt_xband}. Throughout this paper, all magnitudes have been corrected for Galactic extinction. The Galactic extinction values were obtained from NED\footnote{\url{https://ned.ipac.caltech.edu/}} \citep{NED}, with values of 0.378\,mag for the VT$\_R$ band and 0.632\,mag for the VT$\_B$ band.

The MXT observed PKS 1725+123 a total of nine times from 2025 August 20 to 2025 September 08. Intervals affected by stray light, Earth occultations, and passages of the satellite through the South Atlantic Anomaly were excluded, and the net exposure times for each observation are listed in Table~\ref{tab:svom}. The event-mode data were processed with the MXT pipeline v1.13 \citep{maggi2026mxtpipeline}. 

The information regarding the ECLAIRs observations is also presented in Table \ref{tab:svom}. The reduction and analysis of ECLAIRs data for PKS~1725+123 were performed using the ECLAIRs pipeline \citep{goldwurm2026eclairs_pipeline}, which enables both imaging and spectroscopic analysis from event data. As a first step, good time intervals (GTIs) were selected based on satellite attitude stability and instrumental conditions. Calibrated detector-plane images were then generated in a series of energy bins between 4 and 30~keV. Detector images in several broad energy bins were spatially deconvolved using the coded-mask pattern to produce sky images. A point-source extraction algorithm was subsequently applied to the sky images to fit the positions of all sources within the field of view. 

\section{Results} \label{sec:result}

\subsection{SVOM-VT}\label{resultsec:svom-vt}

\begin{figure*}[ht!]
    \centering
    \includegraphics[width=18cm]{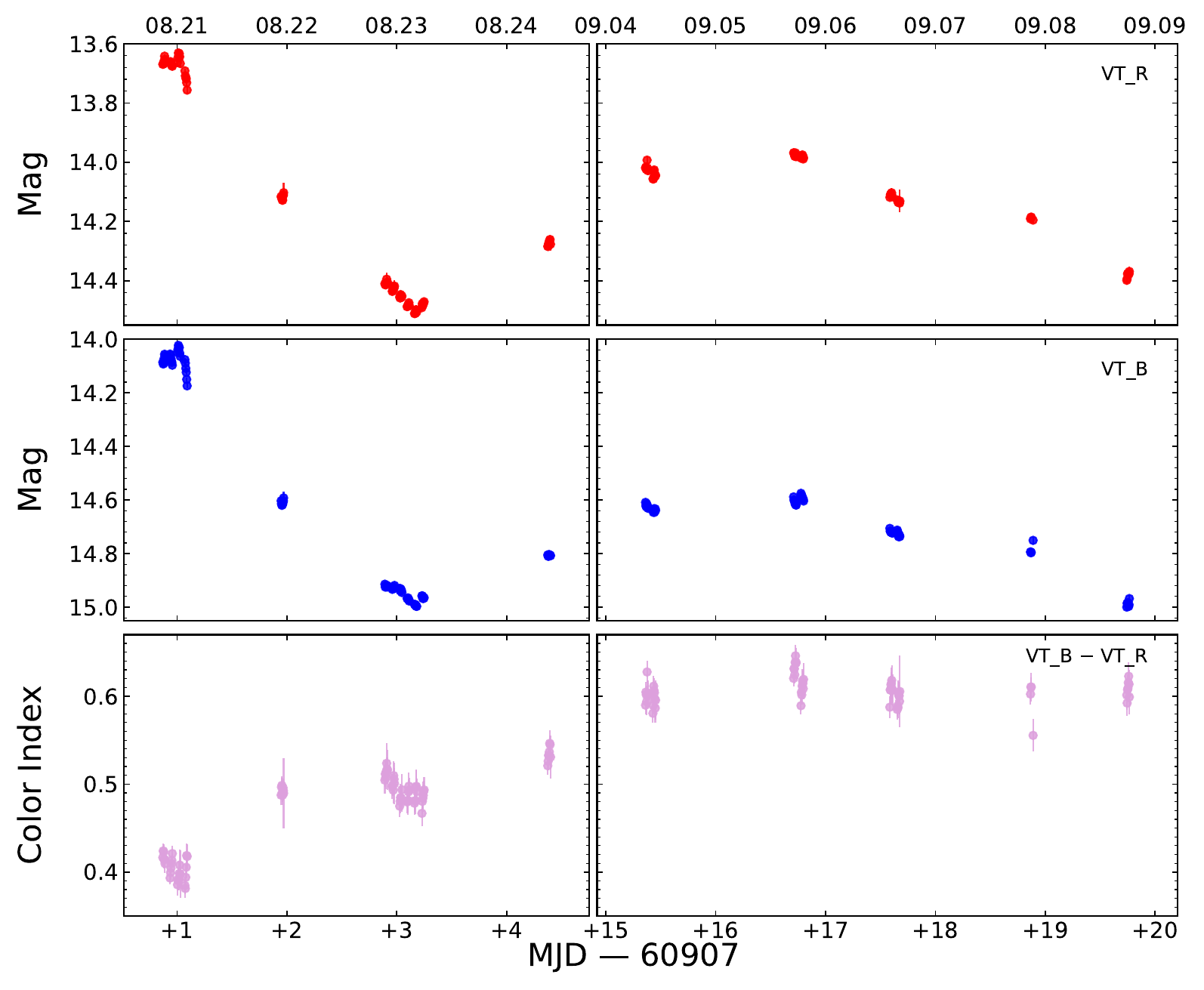}
     \caption{Light curves of magnitudes observed by SVOM-VT in the VT$\_R$ (top panel) and VT$\_B$ (middle panel) bands for PKS 1725+123, along with the color index curve (bottom panel). The data analysis for each orbit observation uses a 5-minute time bin. The first SVOM-VT observation for the source was carried out on 2025 August 20 (MJD 60907).}
      \label{lc-VT_day}
    \end{figure*}

We produced the SVOM-VT light curves of PKS 1725+123 in both the VT$\_R$ and VT$\_B$ bands with a time bin of 5 minutes, as shown in Figure \ref{lc-VT_day}. The source exhibits significant variability on daily timescales during the SVOM-VT observations. The brightness of PKS 1725+123 in both the VT$\_R$ and VT$\_B$ bands declines significantly from 2025 August 20 to August 23, then increases, and subsequently decreases again on 2025 September 5. The brightest magnitude is 13.62$\pm$0.01 mag in the VT$\_R$ band and $14.02\pm0.02$ mag in the VT$\_B$ band observed on 2025 August 20. The brightness variation of the source is approximately 0.89 mag in the VT$\_R$ band and approximately 0.98 mag in the VT$\_B$ band from August 20 to August 23.

To quantify the significance of variability, a weighted mean flux $\langle{F}\rangle$ is estimated using \citep{1996ApJ...473..763M}
\begin{equation}
\langle{F}\rangle=\left[\sum\limits_{i=1}^N\frac{F_i}{\sigma^{2}_i}\right]\left[\sum\limits_{i=1}^N\frac{1}{\sigma^{2}_i}\right]^{-1},
\end{equation}
where $N$ is the number of the data points, ${F_i}$ and $\sigma_{i}$ are the flux and its error for the $i$th data point. We assume a constant flux of $\langle{F}\rangle$ and subsequently calculate the $\chi^2$ value by 
\begin{equation}
\chi^2=\sum\limits_{i=1}^N\frac{(F_i-\langle{F}\rangle)^2}{\sigma_{i}^2},
\end{equation}
and the associated probability $p(\chi^2)=1-p(>\chi^2)$. The results indicate that the significance of the variability in both the VT$\_R$ and VT$\_B$ bands far exceeds the $5\sigma$ confidence level, with the estimated values of $p(>\chi^{2})$ being considerably lower than $5.7\times10^{-7}$.

To investigate whether PKS 1725+123 exhibits significant variability on hourly timescales, we reproduce the VT light curves obtained from several consecutive orbital observations conducted within a 24-hour period. As shown in Figure \ref{lc-VT_hour}, there are two observation epochs: consecutive four-orbit observations from August 20 to August 21, and consecutive six-orbit observations from August 22 to August 23. The data analysis uses a time bin of 5 minutes. Using the Equations (1) and (2), we estimate the significance of variability for these light curves. The results indicate that the source exhibits significant variability on hourly timescales during both observation epochs in both the VT$\_R$ and VT$\_B$ bands, with a confidence level far exceeding $5\sigma$. The weighted mean flux $\langle{F}\rangle$ for each observation epoch is also presented as dashed lines in Figure \ref{lc-VT_hour}.

We further assess the significance of variability for each orbit observation, as presented in Table \ref{tab:svom}. The data analysis also uses a time bin of 5 minutes. The results suggest that the source shows significant variability with a confidence level exceeding $5\sigma$ in the VT$\_B$ band during the second and fourth orbit observations on August 20--21, as displayed in Figure \ref{lc-VT_hour}. In contrast, there is no variability during the second orbit observation and variability with a 3.5$\sigma$ confidence level during the fourth orbit observation in the VT$\_R$ band. Evidently, the variability of PKS 1725+123 in the optical band can be as short as several minutes. Nevertheless, no significant ($>5\sigma$) variability is observed for the remaining orbit observations. 

The time evolution of the color index are also presented in Figures \ref{lc-VT_day} and \ref{lc-VT_hour}, where the color index is defined as the magnitude in the VT$\_B$ band minus that in the VT$\_R$ band. It can be observed that the color index, which accompanies the magnitude variation, also exhibits evident variation on daily timescales (Figure \ref{lc-VT_day}), with a confidence level exceeding 5$\sigma$. Subsequently, we plot the color index versus the magnitudes in both the VT$\_R$ and VT$\_B$ bands for all the observational data, as displayed in Figure \ref{color-index} (a) and (b). It appears that the data are distributed into two different clusters. Considering the uncertainties associated with both the color index and the magnitude, we use the bootstrap method \citep{Efron:1979bxm} to estimate the correlation coefficient ($r$) between the color index and the magnitude. We obtain $r = -0.78\pm0.05$ and $r = 0.10\pm0.17$ when the VT$\_R$ magnitude values are higher (fainter) and lower (brighter) than 13.8 mag, respectively. Similarly, we obtain $r = -0.61\pm0.08$ and $r = 0.45\pm0.11$ when the VT$\_B$ magnitude values are higher (fainter) and lower (brighter) than 14.2 mag, respectively. These results likely indicate a \textit{redder when brighter} (RWB) trend for the source if it is fainter than 13.8 mag in the VT$\_R$ band (or 14.2 mag in the VT$\_B$ band).

Note that the points with magnitudes brighter than 13.8 mag in the VT$\_R$ band (or 14.2 mag in the VT$\_B$ band) in Figure \ref{color-index}(a) (or Figure \ref{color-index}(b)) correspond to the data from the first four orbit observations shown in Figure \ref{lc-VT_day} (and to the left panels in Figure \ref{lc-VT_hour}). Therefore, we further estimate the correlation coefficient between the color index and the magnitude for each of the four orbit observation datasets. The bootstrap analysis yields $r=0.78\pm0.18$ ($r=0.93\pm0.09$ for VT$\_B$) and $r=0.74\pm0.18$ ($r=0.90\pm0.14$ for VT$\_B$) for the VT$\_R$ observation data of the second (Figure \ref{color-index} (c) and (d)) and fourth ( Figure \ref{color-index} (e) and (f)) orbits, whereas no correlation is observed for the remaining two orbit observations. Clearly, the source exhibits a \textit{bluer when brighter} (BWB) behavior during the second and fourth orbital observations. 

\subsection{SVOM-MXT and SVOM-ECLAIRs}

The source was not significantly detected in the SVOM-MXT observations. We thus assume an absorbed power-law (PL) spectrum with a Galactic column density of $N_{\rm H}=8.64\times10^{20}~{\rm cm^{-2}}$ \citep{2016A&A...594A.116H} and a photon index of $\Gamma_{\rm X} = 2.0$ to compute the $3\sigma$ upper limits of the flux in the 0.3--10\,keV energy band. To provide a representative constraint across all pointings, we calculated a weighted-average upper limit of $4.7 \times 10^{-12}\,\mathrm{erg\,cm^{-2}\,s^{-1}}$, where the weighting accounts for the exposure time of each observation.

No significant emission was detected by the SVOM-ECLAIRs during the observational campaign. Therefore, the $3\sigma$ upper limits of the flux were calculated for each individual observation using the variance maps produced by the ECLAIRs pipeline. A $3\sigma$ weighted-average upper limit of the flux in the 4--150\,keV energy band, which accounts for the exposure duration of each pointing, was finally computed as $5.8 \times 10^{-10}\,\mathrm{erg\,cm^{-2}\,s^{-1}}$, assuming an $E^{-2}$ PL spectrum.

\subsection{Swift-XRT}\label{sec_XRT}

We also conduct an analysis of the X-ray observational data of the source from Swift-XRT. The data reduction procedure of the Swift-XRT observations is described in Appendix A.1. We derive the X-ray light curves of PKS 1725+123 in the 0.3--2\,keV ($F_{0.3-2}$), 2--10\,keV ($F_{2-10}$), and 0.3--10\,keV ($F_{0.3-10}$) bands, along with the temporal curves of photon index ($\Gamma_{\rm X}$) and hardness ratio (HR, i.e., $\frac{F_{2-10}-F_{0.3-2}}{F_{2-10}+F_{0.3-2}}$), as displayed in Figure \ref{lc-XRT}. 

The light curves of $F_{0.3-2}$, $F_{2-10}$, and $F_{0.3-10}$ for PKS 1725+123 exhibit significant variability at a confidence level exceeding 5$\sigma$. It should be noted that, based on the Swift-XRT observations, the source on 2025 August 19 and August 21 (two data points within the shaded region) is in a historically high-flux state. This state coincides with the VHE detections and the highest SVOM-VT flux observations.

It is noteworthy that the value of $F_{0.3-10}$ seems to increase from 2025 August 19 to 21 (two data points within the shaded region). In contrast, the spectrum becomes softer. Five days later, on 2025 August 25 and 26, the flux declines to approximately half of that on 2025 August 21, but with a harder spectral index. Notably, for almost all observations, the flux in the 0.3--10\,keV band is predominantly determined by the value of $F_{2-10}$, with the exception of the two observations that are concurrent with the VHE detections, as shown in the second panel from the top of Figure \ref{lc-XRT}. In other words, the observed decrease in flux from 2025 August 21 to August 25, is mainly attributed to the decline in flux in the soft 0.3--2\,kev energy band.

\subsection{Fermi-LAT}\label{sec_LAT}

The FSRQ PKS 1725+123 has been detected in the GeV $\gamma$-ray band by the Fermi-LAT, and it is associated with the $\gamma$-ray source 4FGL J1728.0+1216 \citep{Ballet2024_4FGLDR4}. Therefore, we conduct an analysis of the 17-year Fermi-LAT observational data for this source. The data reduction procedure is described in Appendix A.2. 

The 17-year integrated spectrum of PKS 1725+123 in the 0.1--1000\,GeV band requires a log-parabola (LP) model for adequate explanation, with a photon spectral index of $\Gamma_{\gamma}=2.16\pm0.03$ and a curvature parameter of $\beta=0.05\pm0.01$, as presented in Figure \ref{Sectrum_LAT}. We generate the 17-year Fermi-LAT light curve in the 0.1--1000\,GeV band with a time bin of 30 days, as shown in Figure \ref{lc-LAT}(a) in Appendix A.2. The $\gamma$-ray emission of the source remains nearly in a low and stable flux state prior to MJD 60353 (2025 February 12), after which the $\gamma$-ray flux increases, particularly during 2025 August. To further explore the flux state and the variability features in the GeV band before and after the TeV detection, we produce the light curve with a time bin of one day from 2025 July 06 (MJD 60862) to 2025 October 01 (MJD 60949), as shown in Figure \ref{lc-LAT}(b) in the Appendix A.2. Nevertheless, no significant variability is detected in the three-month light curve despite the source being in a high-flux state.

Using the \textit{gtsrcprob} tool, we estimated the maximum energy of photons detected during the entire 17-year Fermi-LAT observation. The maximum energy of the detected photons is 103 GeV, detected on 2025 August 17 (MJD 60904), with a probability exceeding 99\%. To investigate the spectral variation feature of PKS 1725+123 in the GeV band during the high-flux state in the optical band, we perform a time-resolved spectral analysis for the Fermi-LAT observational data in the following time intervals: 2025 August 16--18 (MJD 60903--60905), August 19--21 (MJD 60906--60908), and August 22--24 (MJD 60909--60911). The results of the spectral analysis are also presented in Figure \ref{Sectrum_LAT}. The three time-resolved spectra can be well modeled by a simple PL function with a flatter photon spectral index than that of the 17-year integrated spectrum. As illustrated in Figure \ref{Sectrum_LAT}, minor flux variations are observed at the high-energy end among the three time-resolved spectra, and the spectrum appears to soften from August 16 to August 24. Specifically, the flux is $(2.45\pm1.77)\times 10^{-9}\,\mathrm{erg\,cm^{-2}\,s^{-1}}$ with $\Gamma_{\gamma}=1.72\pm0.12$ during August 16--18, $(7.47\pm2.97)\times 10^{-10}\,\mathrm{erg\,cm^{-2}\,s^{-1}}$ with $\Gamma_{\gamma}=1.98\pm0.14$ during August 19--21, and $(6.10\pm1.77)\times 10^{-10}\,\mathrm{erg\,cm^{-2}\,s^{-1}}$ with $\Gamma_{\gamma}=2.15\pm0.50$ during August 22--24. It seems that PKS 1725+123 demonstrates a \textit{harder-when-brighter} trend in the GeV band.

\section{SED Constructing and Modeling}\label{sec:SED}

As stated in Section \ref{sec_LAT}, a hard Fermi-LAT spectrum is observed for PKS 1725+123 during the TeV detection. Moreover, a high-energy photon of 103 GeV is detected on 2025 August 17 (MJD 60904). We present the sensitivity curves of the MAGIC telescopes (50 hr) and the High Energy Stereoscopic System (H.E.S.S., 50 hr) in Figure \ref{Sectrum_LAT}. The sensitivity curves are taken from \citet{2016APh....72...76A} and \citet{2015ICRC...34..980H}, respectively. The high-energy end of the time-resolved spectrum during August 16--18 exceeds the sensitivity curves of the MAGIC telescopes and H.E.S.S.. To further investigate the $\gamma$-ray emission property and radiation mechanisms of PKS 1725+123 in Multiwavelengths, we construct its broadband SED in the high-flux state, as illustrated in Figure \ref{SED}. The average flux in both the VT$\_R$ and VT$\_B$ bands, observed by SVOM-VT on 2025 August 20--21 (the first consecutive four-orbit observations), the average spectrum of two Swift-XRT observations on 2025 August 19 and 21, and the time-resolved Fermi-LAT spectrum in 2025 August 16--18, are considered as the data for the high-flux state. The data in the low-energy band from the NED, the average spectrum of the first three Swift-XRT observations (left panels in Figure \ref{lc-XRT}), and the 17-year Fermi-LAT average spectrum are also presented in Figure \ref{SED} for comparison. It should be noted that the broadband SED of the source in the high-flux state exhibits significant differences from that of the average-flux state (gray symbols in Figure \ref{SED}), showing obvious variability and spectral variation in Multiwavelengths.   

A two-zone leptonic model is used to reproduce the broadband SED of PKS 1725+123. This model bears resemblance to the morphology model of Mrk 421 presented by \citet{2025A&A...695A.217M}, in which the Doppler factor ($\delta$) for both the extended and compact zones is fixed at 60. Nevertheless, it more closely resembles a structured spine-sheath jet model (e.g., \citealt{2005A&A...432..401G, 2008MNRAS.385L..98T}). Specifically, a compact zone with a large bulk Lorentz factor ($\Gamma$) is situated at the center of an extended region with a small value of $\Gamma$. Due to plasma instability, magnetic reconnection may occur to accelerate the electrons within the compact zone (e.g., \citealt{2011MNRAS.413..333N, 2013MNRAS.431..355G, 2016MNRAS.462.3325P}), leading to high flux with rapid variability. Meanwhile, the relativistic electrons within the compact region propagate outward and traverse the jet shear layer, which is the transitional zone between the spine and the sheath. Here, the relativistic electrons are accelerated once again by the shear acceleration process (e.g., \citealt{2017ApJ...842...39L, 2019Galax...7...78R, 2021ApJ...907L..44S}). It is assumed that both zones are located within the broad line regions (BLRs), yet they are in proximity to the outer boundary of the BLRs. Therefore, the synchrotron, synchrotron self-Compton (SSC), and external Compton (EC) scattering of the photons from BLR (EC/BLR) processes of the relativistic electrons in both regions are taken into account to reproduce the broadband SED of the source. Given that the two regions overlap, the interaction between the two zones results in additional emission as they provide an extra target photon field for IC scattering for each other, and the contribution of which is also considered. For more details regarding the model assumptions and parameter sets, please refer to Appendix B.1.

The fitting result is presented in Figure \ref{SED}. The synchrotron radiation of relativistic electrons within the compact zone produces the optical--X-ray emission, which is also consistent with the short timescale variability observed by SVOM-VT and the absence of high-energy photons during the Swift-XRT observations. The low-energy end of the Fermi-LAT spectrum is attributed to the SSC+EC/BLR processes of this electron population, whereas the high-energy end of the Fermi-LAT spectrum results from the IC scattering of synchrotron photons from the compact zone by relativistic electrons within the extended region. It can be observed that the intrinsic spectrum in the $\gamma$-ray band is very hard (represented by the green dash-dotted line) and significantly surpasses the sensitivity curves of the H.E.S.S. and MAGIC telescopes. After accounting for the extragalactic background light (EBL) absorption \citep{2022ApJ...941...33F}, the model-predicted flux (represented by the black solid line) at the VHE band still exceeds the sensitivity curves of the H.E.S.S. and MAGIC telescopes, which is consistent with the MAGIC and H.E.S.S. detections \citep{2025ATel17344....1P,2025ATel17346....1W}. Additionally, the SVOM-VT data, in combination with the Swift-XRT spectrum, suggests a very high synchrotron peak frequency for PKS 1725+123, which is distinctly different from that of previously $\gamma$-ray-detected FSRQs (e.g., \citealt{2014ApJ...788..104Z, 2015ApJ...807...51Z, 2015MNRAS.452.1303P, 2025ApJ...989..125M, 2025PASA...42...53M}).

\section{Discussion and Conclusions}\label{sec:dis&con}

It was reported that FSRQ PKS 1725+123 was detected for the first time in the VHE band by the MAGIC and LST-1 telescopes on the night of 2025 August 18--19 \citep{2025ATel17344....1P}, and by H.E.S.S. on the night of 2025 August 19--20 \citep{2025ATel17346....1W}. Following these reports, SVOM performed a ToO monitoring campaign of the source from 2025 August 20 to 2025 September 8. SVOM-VT observations indicate that the source brightness reaches 13.62$\pm$0.01 mag in the VT$\_R$ band and $14.02\pm0.02$ mag in the VT$\_B$ band on 2025 August 20. This may represent a historical optical light maximum and is brighter than the value of $R=14.21\pm0.02$ mag (perhaps without accounting for Galactic extinction) on 2025 August 19, reported by \cite{2025ATel17345....1G}. Moreover, significant variability is observed on a timescale of less than an hour, on the order of several minutes. Considering the rapid and intense variability, along with the follow-up very high optical polarization measurement ($\sim$35\% in the $R$-band) on 2025 September 15 \citep{2025ATel17393....1B}, the observed optical flux of the source should be dominated by jet radiation rather than its host galaxy.

\begin{figure*}[ht!]
    \centering
    \includegraphics[width=18cm]{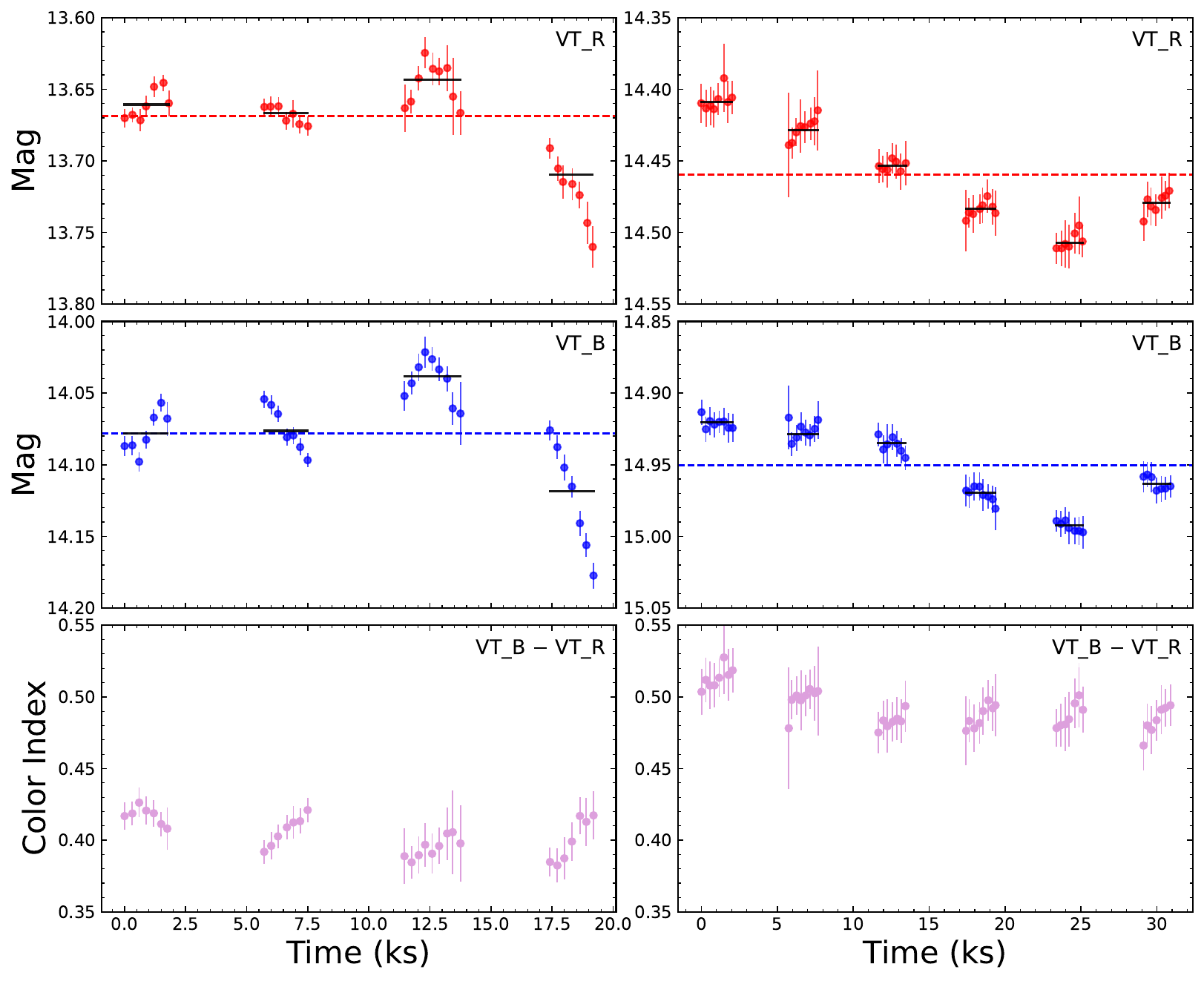}
     \caption{The partial zoom-out of Figure \ref{lc-VT_day}. Only the data from the consecutive four-orbit observations conducted on 2025 August 20--21 (left panels) and the consecutive six-orbit observations conducted on 2025 August 22--23 (right panels) are presented. The time bin of the data points is also 5 minutes, the same as in Figure \ref{lc-VT_day}. The horizontal dashed lines indicate the weighted-mean magnitude for the consecutive four-orbit (and six-orbit) observations, derived using Equation (1). The horizontal black solid lines denote the weighted-mean magnitude for each individual orbit observation.}
      \label{lc-VT_hour}
    \end{figure*}

\begin{figure*}[ht!]
    \centering
    \includegraphics[width=8cm]{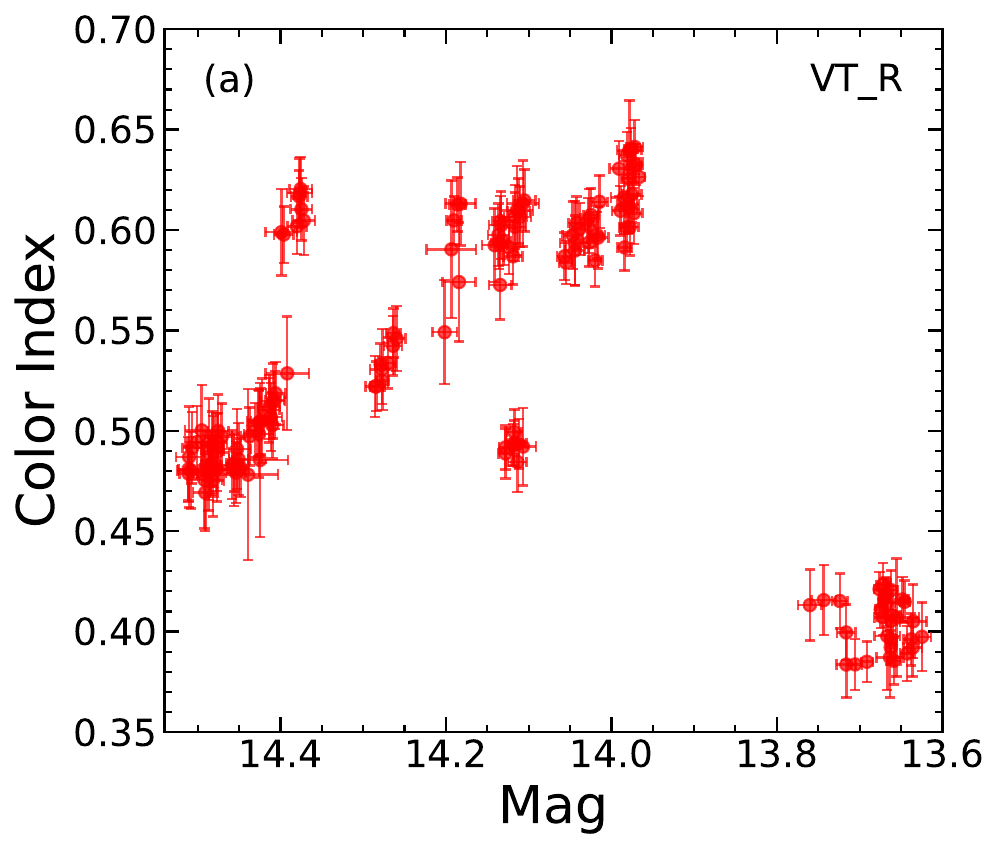}
    \includegraphics[width=8cm]{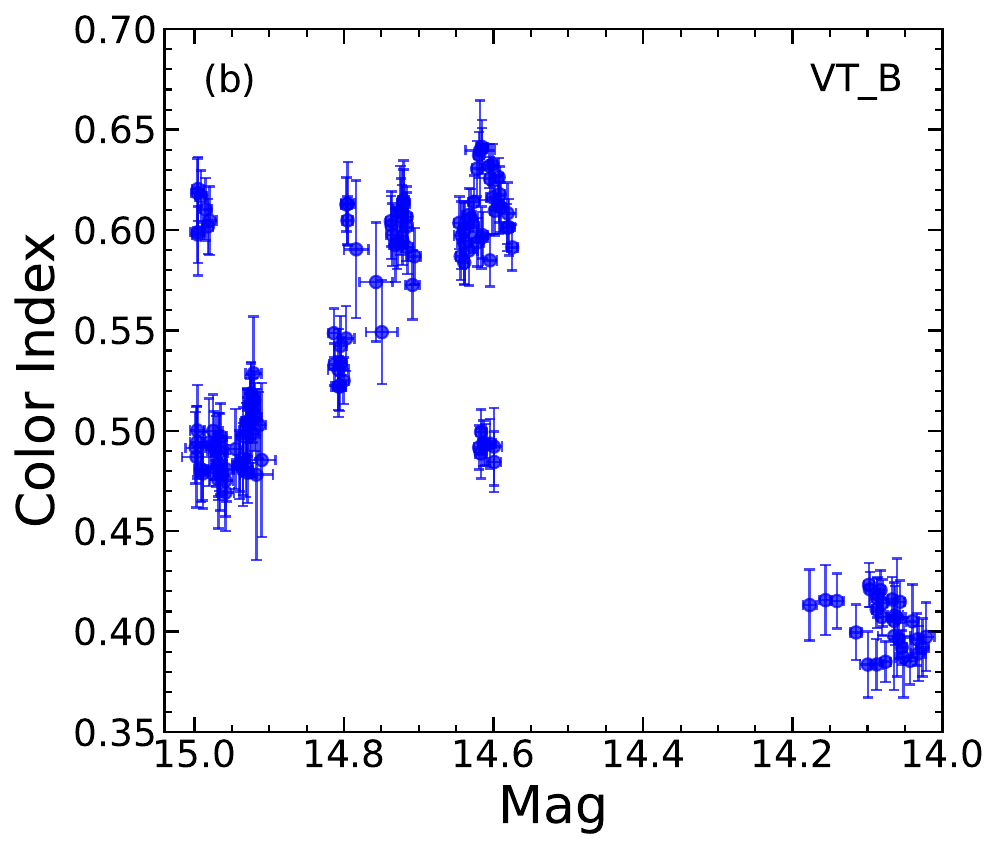}\\   
    \includegraphics[width=8cm]{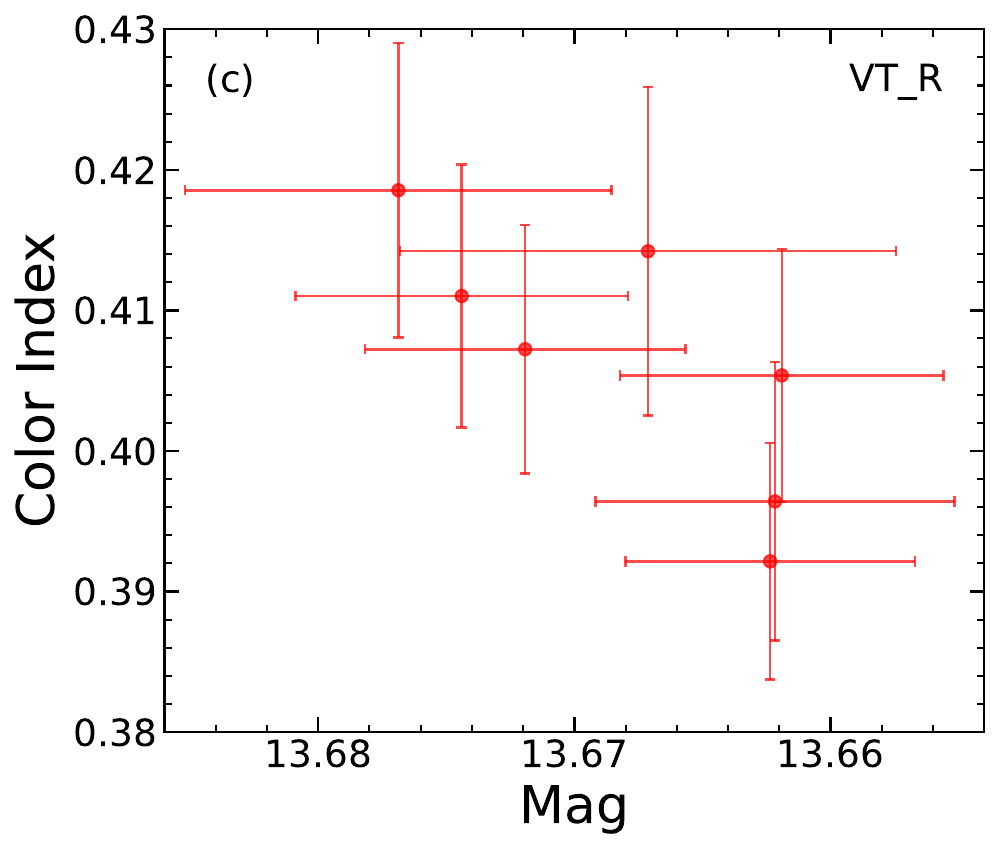}
    \includegraphics[width=8cm]{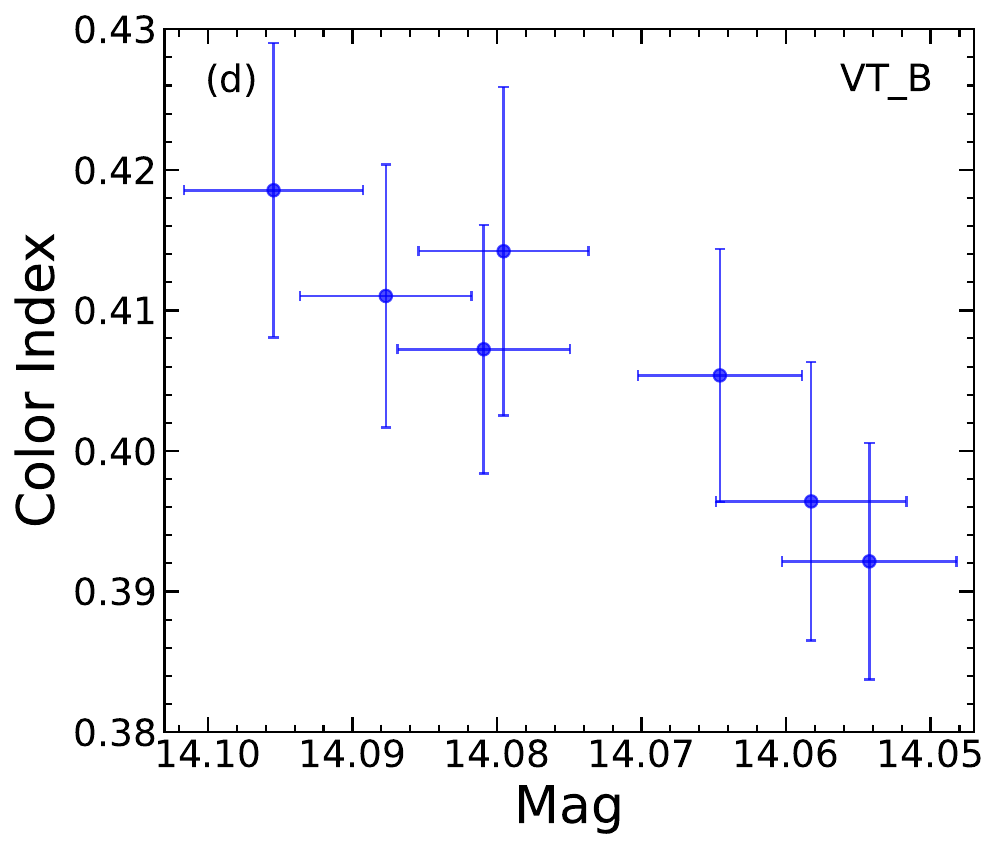}\\
    \includegraphics[width=8cm]{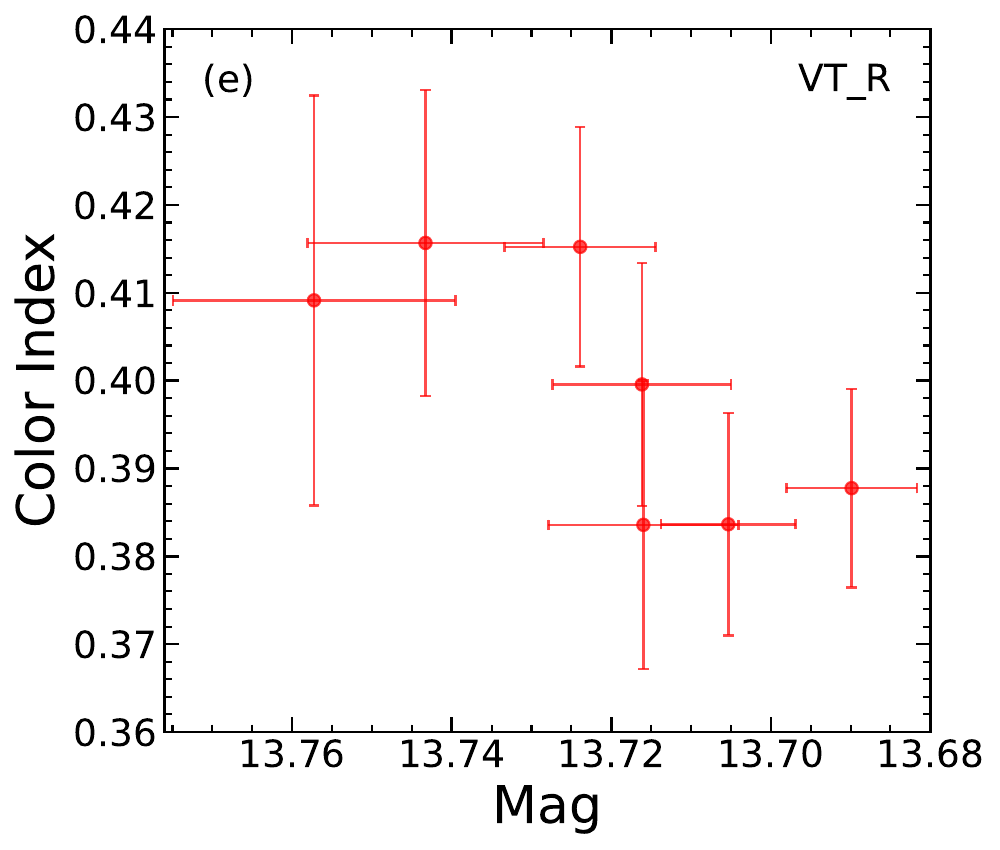}
    \includegraphics[width=8cm]{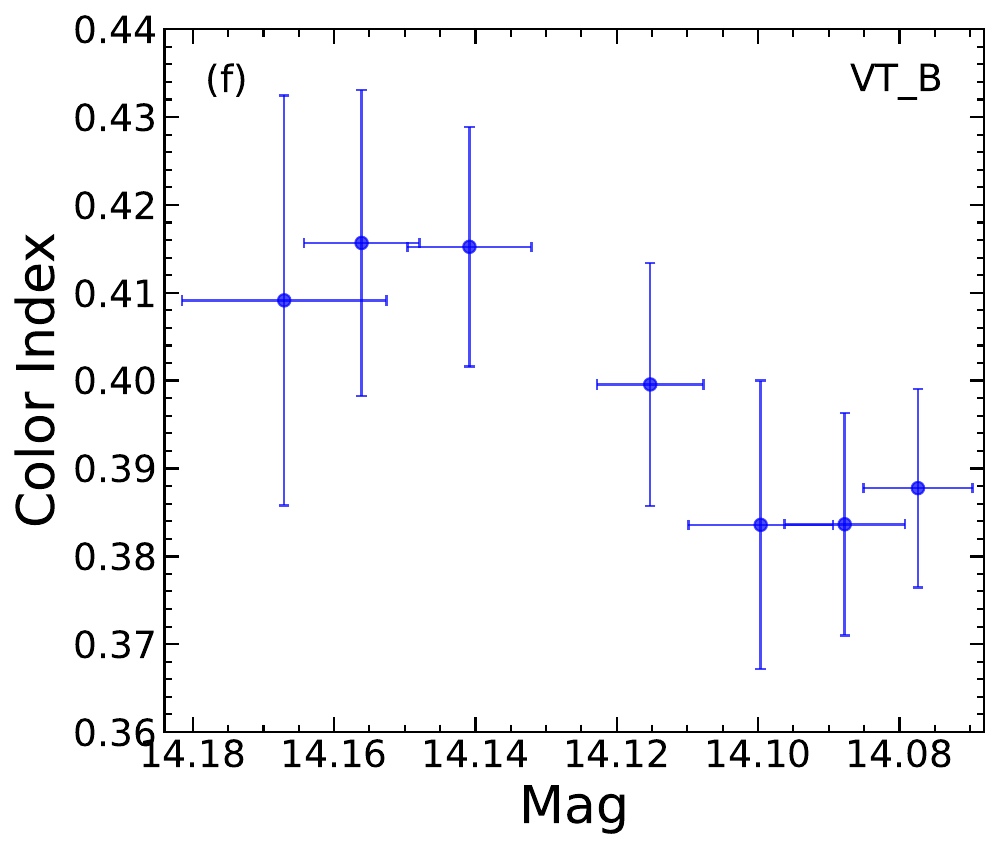}
     \caption{Color index vs. magnitudes in both the VT$\_R$ (left panels) and VT$\_B$ (right panels) bands for PKS 1725+123. Panels (a) and (b) show all the SVOM-VT observational data, identical to those in Figure \ref{lc-VT_day}. Panels (c) and (d) present the observational data from the second orbit of the consecutive four-orbit observations conducted on 2025 August 20--21. Panels (e) and (f) illustrate the observational data from the fourth orbit of the consecutive four-orbit observations conducted on 2025 August 20--21.}
      \label{color-index}
    \end{figure*}

As presented in Figure \ref{color-index}, the variation in optical brightness is accompanied by the change of the color index. Both the RWB and BWB trends are a prevalent phenomenon in some FSRQs (e.g., \citealt{2006A&A...453..817V,2017Natur.552..374R,2025A&A...703A.259V}). In low-flux state, sources generally display the RWB trend, which is interpreted as the increasing relative contribution from the accretion disk thermal emission when the jet radiation fades. In contrast, when the sources are in high-flux states, they exhibit an obvious BWB trend, which is also a commonly observed behavior in numerous BL Lac objects \citep{2015A&A...573A..69W}. This phenomenon is typically attributed to the dominant emission from the relativistic jets. Fundamentally, it may be related to the injection or re-acceleration of electrons within the jets (e.g., \citealt{1998A&A...333..452K,2001ApJ...554....1S,2002PASA...19..138M,2025A&A...703A.259V}). The FSRQ PKS 1725+123 exhibits strongly chromatic short-timescale changes and significant BWB behavior in the high-flux state during the SVOM-VT monitoring, as present in Figures \ref{lc-VT_hour} and \ref{color-index}. This phenomenon is due to energetic processes occurring in the jet (e.g., \citealt{2025A&A...703A.259V}).  

\begin{figure*}[ht!]
    \centering
    \includegraphics[width=0.9\textwidth]{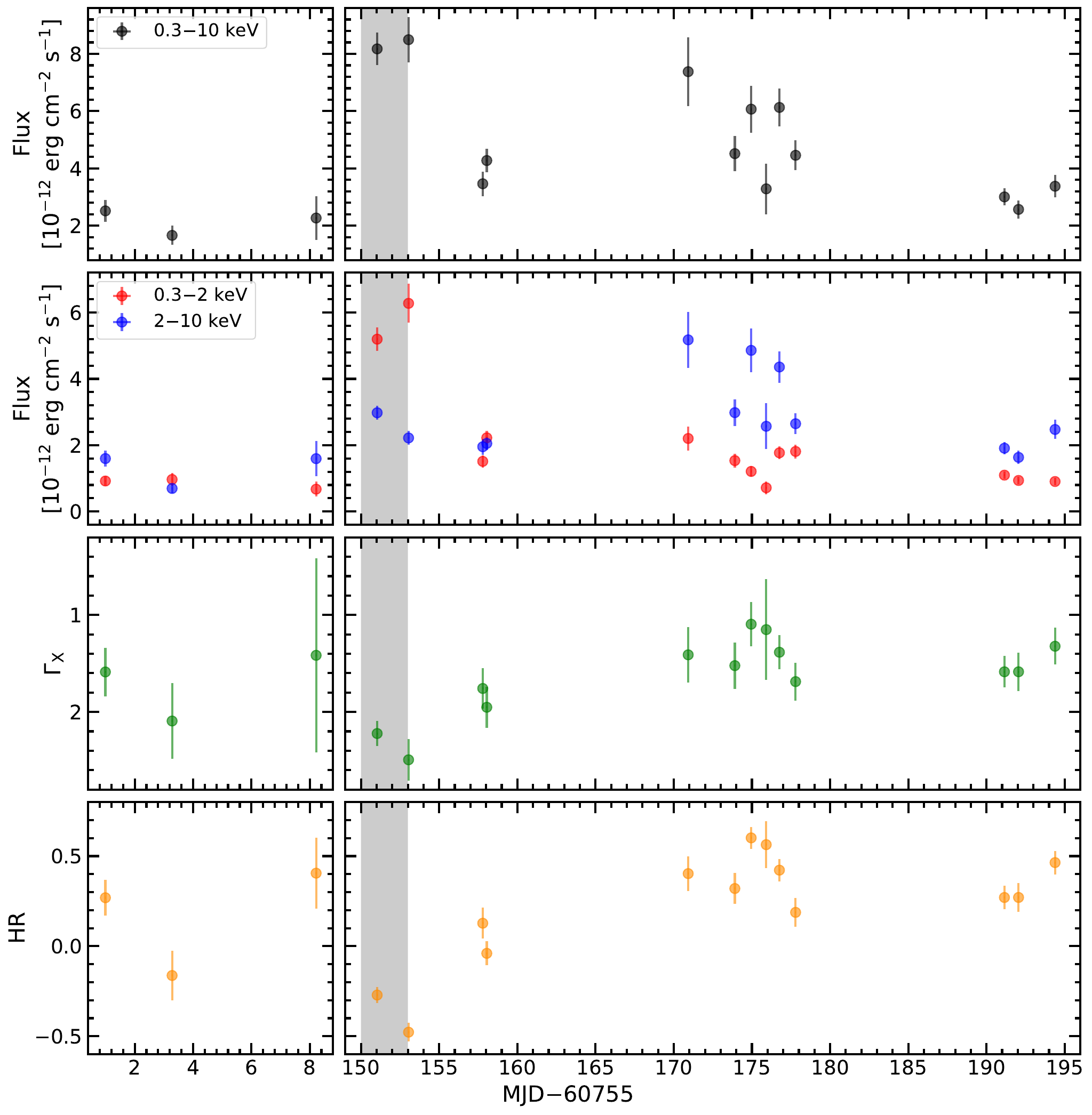}
    \caption{Swift-XRT observations for PKS 1725+123. Light curves of $F_{0.3-10}$ (black points in the top panels), $F_{0.3-2}$ (red points in the second panels), $F_{2-10}$ (blue points in the second panels), $\Gamma_{\gamma}$ (green points in the third panels), and the HR (orange points in bottom panels) are respectively depicted. The gray shaded area indicates the quasi-simultaneous observations (on August 19 and 21) corresponding to the TeV detections. The left panels show the three historical Swift-XRT observations, while the right panels display the Swift-XRT follow-up observations of the TeV detection for the source. MJD 60755 corresponds to 2025 March 21.}
    \label{lc-XRT}
\end{figure*}

Based on Swift-XRT observations, the X-ray emission of the source is also in a high-flux state (Figure \ref{lc-XRT}) during its optical high-flux state. However, different from the chromatic behavior of BWB in the optical band, the simultaneously observed X-ray emission of PKS 1725+123 displays a \textit{softer when brighter} trend, as shown in Figure \ref{lc-XRT}. This characteristic is also entirely different from the observed X-ray features in high-synchrotron-peaked BL Lac objects. These sources generally exhibit the \textit{harder when brighter} behavior in the X-ray band, such as Mrk 421 \citep{2009A&A...501..879T,2022ApJ...938L...7D}, PKS 2155--304 \citep{2024ApJ...963L..41H}, Mrk 501 \citep{2024ApJ...970L..22H}, H 1426+428 \citep{2025ApJ...986..182H}, which is usually interpreted as the injection of high-energy electrons into the emission region \citep{1998A&A...333..452K,2022ApJ...938L...7D,2024ApJ...965...58Z}. Within the framework of the leptonic jet radiation model, the X-ray emission of FSRQs is generally ascribed to the IC scattering process (e.g., \citealt{2014ApJ...788..104Z,2015ApJ...807...51Z}). It should be noted that the X-ray emission from PKS 1725+123 observed on 2025 August 19--21 is in a high-flux state with a soft spectrum, as illustrated in Figure \ref{lc-XRT}. The soft X-ray spectrum of the source may be due to the synchrotron tail extending into the soft X-ray band during a flare, resulting in a softer spectrum as the source brightens. 

\begin{figure*}[ht!]
    \centering
    \includegraphics[width=1.0\textwidth]{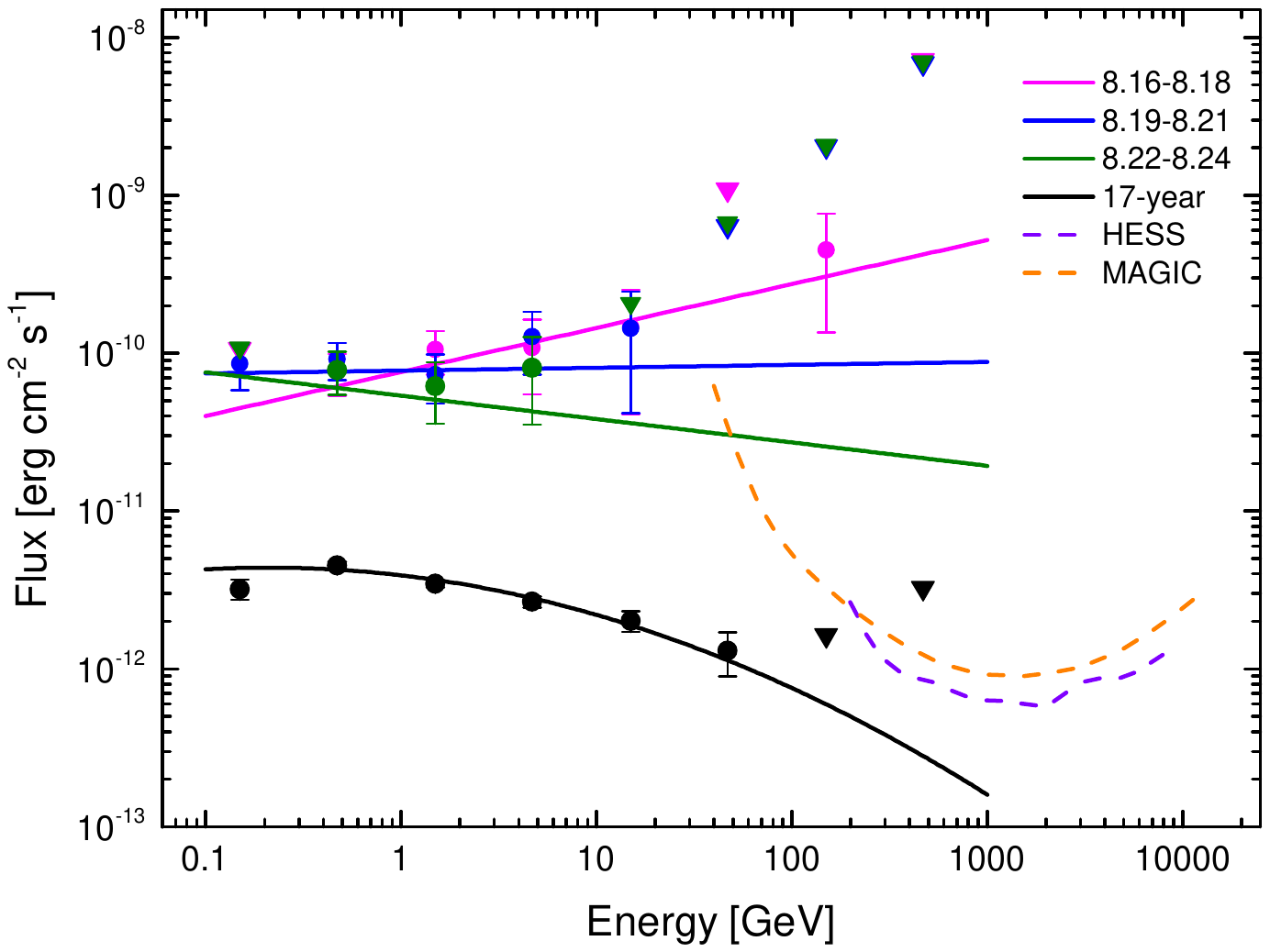}
    \caption{Spectra observed by Fermi-LAT for PKS 1725+123, including the 17-year integrated spectrum (black symbols) and the three time-resolved spectra in the following time intervals: 2025 August 16--18 (magenta symbols), August 19--21 (blue symbols), and August 22--24 (green symbols). The corresponding colored solid lines represent the fitting results. If TS$<$4, an upper limit (denoted by inverted triangles) is given for that energy bin. The sensitivity curves of the MAGIC telescopes (orange dashed line, 50 hr) and H.E.S.S. (purple dashed line, 50 hr) are also given in the figure.} 
      \label{Sectrum_LAT}
\end{figure*}

Although the $\gamma$-ray emission of PKS 1725+123 is in a high-flux state during 2025 August, no significant variability is observed on a daily timescale, as depicted in Figure \ref{lc-LAT}(b) in Appendix A.2. A high-energy photon of 103 GeV is detected on 2025 August 17 by the Fermi-LAT. Given the limited sensitivity to resolve the spectrum of PKS 1725+123 below daily timescales and the low variability in this waveband, we generated the 3-day interval spectra around the TeV detection time, including three time intervals: 2025 August 16--18 (MJD 60903--60905), August 19--21 (MJD 60906--60908), and August 22--24 (MJD 60909--60911), as presented in Figure \ref{Sectrum_LAT}. A \textit{harder-when-brighter} trend in the GeV band is displayed during the observations of the three time intervals. This spectral evolution characteristic in the $\gamma$-ray band has been observed in some FSRQs (e.g., \citealt{2010ApJ...721.1425A,2018RAA....18...40Z,2020PASJ...72...44Z}). According to the fitting results of the three time-resolved spectra, we extrapolate the flux into the VHE band, which exceeds the sensitivity curves of the H.E.S.S. and MAGIC telescopes. This is consistent with the reported TeV detections \citep{2025ATel17344....1P,2025ATel17346....1W}.

Although we are unable to quantify the potential correlation of variability among the optical, X-ray, and GeV $\gamma$-ray bands due to the lack of simultaneous and continue monitoring data in Multiwavelengths, these results suggest that the radiations of PKS 1725+123 in the optical, X-ray, and GeV $\gamma$-ray bands are likely correlated. Based on these factors, we construct the broadband SED of PKS 1725+123 in the high-flux state and model it using a two-zone spine-sheath jet model. We hypothesize that the electrons may be accelerated by magnetic reconnection within a compact zone (e.g., \citealt{2011MNRAS.413..333N, 2013MNRAS.431..355G, 2016MNRAS.462.3325P}), resulting in the high flux and rapid variability observed in the optical band. This is also consistent with the flat electron spectral index of $p=1.4$ and the high ratio of $U_B/U_{\rm e}$, which are derived by SED fitting (Table \ref{Tab: Parameters} in Appendix B.1). 

Meanwhile, the relativistic electrons within the compact region propagate outward and traverse the jet shear layer, where they are effectively accelerated again via the shear acceleration process (e.g., \citealt{2017ApJ...842...39L, 2019Galax...7...78R, 2021ApJ...907L..44S}). In this framework, the synchrotron radiation of the relativistic electrons within the compact zone produces the optical--X-ray emission. The IC scattering processes of this electron population contribute to the low-energy end of the Fermi-LAT spectrum, while the high-energy end of the Fermi-LAT spectrum is attributed to the IC scattering of the synchrotron photons within the compact zone by the higher-energy electrons in the extended region. 

\begin{figure*}[ht!]
    \centering
    \includegraphics[width=1\textwidth]{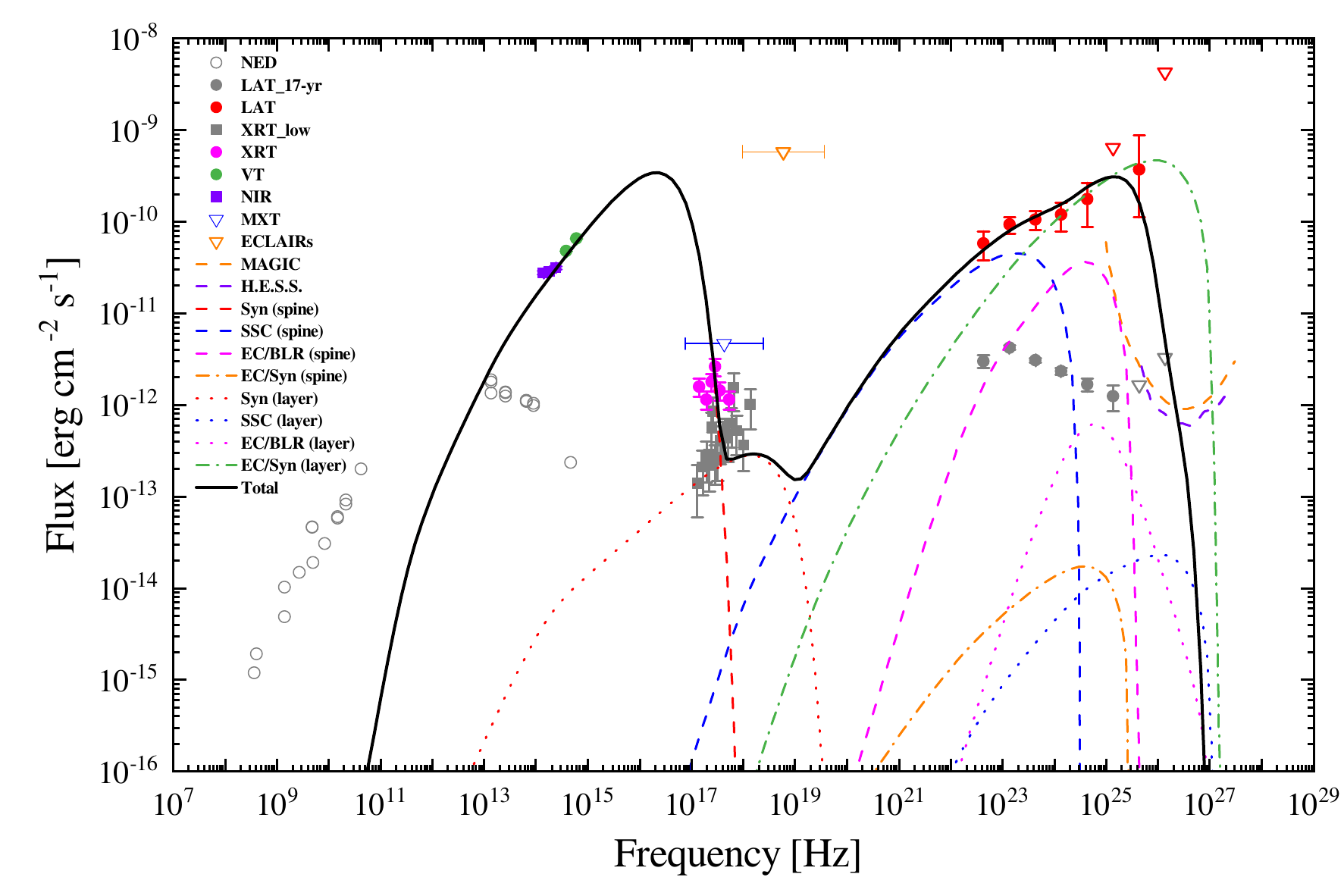}
     \caption{Observed SED with model fitting for PKS 1725+123 in the high-flux state. The three near-infrared (NIR) points (purple solid squares) taken from \citet{2025ATel17415....1C}, the average flux (two green solid circles) from the consecutive four-orbit SVOM-VT observations on 2025 August 20--21, the average spectrum from two Swift-XRT observations (magenta solid circles) on 2025 August 19 and 21 (shaded region in Figure \ref{lc-XRT}), the weighted-average $3\sigma$ upper limit of the flux observed by SVOM-MXT (blue open inverted triangle) and SVOM-ECLAIRs (orange open inverted triangle), and the Fermi-LAT time-resolved spectrum (red solid circles and open inverted triangle) from 2025 August 16--18, collectively constitute the SED of the source in the high-flux state. The black solid line (\textbf{including EBL absorption}) represents the total emission model, including synchrotron radiation (red lines), SSC process (blue lines), and EC/BLR process (magenta lines), where the dashed and dotted lines denote emission from the spine and sheath regions, respectively. The green dash-dotted line indicates the inverse Compton component (\textbf{without EBL absorption}) of the synchrotron photons in the spine zone by the relativistic electrons in the sheath region, while the flux contribution from the opposite process is depicted as the orange dash-dotted line. For comparison, the NED archived radio--optical data (gray open circles), the average spectrum (gray solid squares) from the first three Swift-XRT observations (left panels in Figure \ref{lc-XRT}), and the 17-year Fermi-LAT integrated spectrum (gray solid circles and open inverted triangles) are presented. The sensitivity curves of the MAGIC telescopes (orange dashed line, 50 hr) and H.E.S.S. (purple dashed line, 50 hr) are also given in the figure. }
      \label{SED}
\end{figure*}

As depicted in Figure \ref{SED}, the FSRQ PKS 1725+123 in the high-flux state demonstrates a remarkably high synchrotron peak frequency, which is notably distinct from other $\gamma$-ray emitting FSRQs. In contrast, its broadband SED in the average-flux state (gray symbols in Figure \ref{SED}) is similar to that of typical $\gamma$-ray emitting FSRQs (e.g., \citealt{2014ApJ...788..104Z, 2015ApJ...807...51Z, 2015MNRAS.452.1303P, 2025ApJ...989..125M, 2025PASA...42...53M}). To further investigate the jet properties of PKS 1725+123, we calculate its jet powers and the powers of each jet component for both the compact and extended regions based on the SED fitting parameters (for more details, refer to Appendix B.3). We compare the jet powers and the powers of each jet component of PKS 1725+123 with a $\gamma$-ray emission FSRQ sample from \cite{2008MNRAS.385..283C} and \cite{2020ApJ...899....2Z}, as illustrated in Figure \ref{pjet} in Appendix B. On average, PKS 1725+123 exhibits a low jet power and radiation efficiency, but a highly magnetized jet when compared with other $\gamma$-ray emission FSRQs. Evidently, the simultaneous Multiwavelength observations during the high-flux state of the source offer more constraints for radiation mechanism models and, concurrently, pose new challenges to the traditional radiation models. 

\begin{acknowledgments}
The Space-based multi-band astronomical Variable Objects Monitor (SVOM) is a joint Chinese-French mission led by the Chinese National Space Administration (CNSA), the French Space Agency (CNES), and the Chinese Academy of Sciences (CAS). We gratefully acknowledge the unwavering support of NSSC, IAMCAS, XIOPM, NAOC, IHEP, CNES, CEA, and CNRS. This work is supported by the National Key R\&D Program of China (grants 2024YFA161171 and 2024YFA1611700) and the National Natural Science Foundation of China (grants 12022305 and 11973050).
\end{acknowledgments}

\bibliography{ref}{}
\bibliographystyle{aasjournal}


\begin{table*}
    \begin{center}
    \caption{Results of the SVOM Observations for PKS 1725+123}
    \label{tab:svom}
    \resizebox{\textwidth}{!}{
    \begin{tabular}{ccccccccc}
    \hline
    \hline
    \multicolumn{9}{c}{\bf VT} \\
    \hline
    Orbits & Start Time & End Time & \multicolumn{2}{c}{Magnitude$^{a}$} & \multicolumn{2}{c}{$\sigma^{b}$} & \multicolumn{2}{c}{Details$^{c}$} \\
    \cmidrule(r){4-5} \cmidrule(r){6-7} \cmidrule(r){8-9}
    & (YYYY-MM-DD hh:mm) & (YYYY-MM-DD hh:mm) & R & B & R & B & R & B \\
    \hline
    1  & 2025-08-20 20:22 & 2025-08-20 21:29 & $13.661\pm0.003$ & $14.077\pm0.003$ & 2.4 & 3.9 & 7/29 & 7/28 \\
    2  & 2025-08-20 22:04 & 2025-08-20 23:06 & $13.668\pm0.003$ & $14.076\pm0.002$ & $<2$ & 5.6 & 7/30 & 7/30 \\
    3  & 2025-08-20 23:41 & 2025-08-21 00:43 & $13.645\pm0.004$ & $14.038\pm0.003$ & $<2$ & $<2$ & 9/31 & 9/31 \\
    4  & 2025-08-21 01:19 & 2025-08-21 02:20 & $13.713\pm0.004$ & $14.119\pm0.003$ & 3.7 & 10.1 & 7/30 & 7/30 \\
    5  & 2025-08-21 22:14 & 2025-08-21 23:22 & $14.119\pm0.003$ & $14.612\pm0.003$ & $<2$ & $<2$ & 8/43 & 8/41 \\
    6  & 2025-08-22 21:00 & 2025-08-22 22:01 & $14.409\pm0.005$ & $14.921\pm0.003$ & $<2$ & $<2$ & 8/31 & 8/31 \\
    7  & 2025-08-22 22:36 & 2025-08-22 23:38 & $14.428\pm0.005$ & $14.928\pm0.003$ & $<2$ & $<2$ & 8/29 & 8/29 \\
    8  & 2025-08-23 00:13 & 2025-08-23 01:15 & $14.453\pm0.004$ & $14.936\pm0.003$ & $<2$ & $<2$ & 7/29 & 7/28 \\
    9  & 2025-08-23 01:59 & 2025-08-23 02:52 & $14.483\pm0.004$ & $14.970\pm0.004$ & $<2$ & $<2$ & 8/28 & 8/27 \\
    10 & 2025-08-23 03:28 & 2025-08-23 04:28 & $14.507\pm0.005$ & $14.993\pm0.004$ & $<2$ & $<2$ & 7/26 & 7/27 \\
    11 & 2025-08-23 05:05 & 2025-08-23 06:05 & $14.479\pm0.005$ & $14.963\pm0.003$ & $<2$ & $<2$ & 7/30 & 7/29 \\
    12 & 2025-08-24 08:39 & 2025-08-24 09:36 & $14.274\pm0.003$ & $14.807\pm0.003$ & $<2$ & $<2$ & 9/41 & 9/41 \\
    13 & 2025-09-04 08:15 & 2025-09-04 09:18 & $14.019\pm0.003$ & $14.620\pm0.003$ & $<2$ & $<2$ & 8/41 & 8/37 \\
    14 & 2025-09-04 09:54 & 2025-09-04 10:56 & $14.044\pm0.004$ & $14.640\pm0.003$ & $<2$ & $<2$ & 9/44 & 8/41 \\
    15 & 2025-09-05 16:37 & 2025-09-05 17:39 & $13.974\pm0.003$ & $14.606\pm0.002$ & $<2$ & 2.6 & 9/43 & 9/44 \\
    16 & 2025-09-05 18:15 & 2025-09-05 19:17 & $13.982\pm0.004$ & $14.589\pm0.002$ & $<2$ & $<2$ & 8/45 & 8/44 \\
    17 & 2025-09-06 13:41 & 2025-09-06 14:42 & $14.112\pm0.005$ & $14.719\pm0.003$ & $<2$ & $<2$ & 8/40 & 8/41 \\
    18 & 2025-09-06 15:17 & 2025-09-06 16:19 & $14.133\pm0.004$ & $14.728\pm0.003$ & $<2$ & $<2$ & 8/42 & 8/42 \\
    19 & 2025-09-07 21:18 & 2025-09-07 22:02 & $14.190\pm0.012$ & $14.751\pm0.014$ & $<2$ & $<2$ & 6/21 & 5/18 \\
    20 & 2025-09-08 17:26 & 2025-09-08 18:27 & $14.380\pm0.005$ & $14.990\pm0.003$ & $<2$ & $<2$ & 8/44 & 8/44 \\
    1--20$^{d}$ & 2025-08-20 20:22 & 2025-09-08 18:27 & $14.040\pm0.001$ & $14.607\pm0.001$ & 219.8 & 282.3 & -- & -- \\
    1--4$^{d}$  & 2025-08-20 20:22 & 2025-08-21 02:20 & $13.669\pm0.001$ & $14.078\pm0.001$ & 12.5 & 20.6 & -- & -- \\
    6--11$^{d}$ & 2025-08-22 21:00 & 2025-08-23 06:05 & $14.460\pm0.002$ & $14.950\pm0.001$ & 13.1 & 14.3 & -- & -- \\
    \hline
    \multicolumn{9}{c}{\bf MXT~\&~ECLAIRs} \\
    \hline
    ObsID & Start Time & End Time & \multicolumn{2}{c}{Exposure} & \multicolumn{2}{c}{Flux} & \multicolumn{2}{c}{} \\
    \cmidrule(r){4-5} \cmidrule(r){6-7}
    & (YYYY-MM-DD hh:mm) & (YYYY-MM-DD hh:mm) & MXT & ECLAIRs & MXT$^{e}$ & ECLAIRs$^{f}$ & & \\
    & & & \multicolumn{2}{c}{(s)} & \multicolumn{2}{c}{($10^{-12}~{\rm erg~cm^{-2}~s^{-1}}$)} & & \\
    \hline
    1426069493 & 2025-08-20 20:50 & 2025-08-21 02:56 & 9443  & 2687 & $<3.57$ & $<780.7$  & & \\
    1140856858 & 2025-08-21 22:21 & 2025-08-21 23:58 & 2353  & 921  & $<7.20$ & $<1155.9$ & & \\
    1426069534 & 2025-08-22 21:01 & 2025-08-23 06:43 & 13914 & 4723 & $<2.82$ & $<410.5$  & & \\
    1140856867 & 2025-08-24 08:37 & 2025-08-24 10:15 & 2337  & 1321 & $<7.29$ & $<779.2$  & & \\
    1140856959 & 2025-09-04 08:18 & 2025-09-04 11:32 & 4741  & 2235 & $<5.48$ & $<570.5$  & & \\
    1140856964 & 2025-09-05 16:39 & 2025-09-05 19:53 & 4752  & 4922 & $<5.26$ & $<450.3$  & & \\
    1140856967 & 2025-09-06 13:41 & 2025-09-06 16:55 & 4753  & 4225 & $<5.91$ & $<392.2$  & & \\
    1140856972 & 2025-09-07 20:26 & 2025-09-07 22:03 & 2375  & 1046 & $<7.28$ & $<1800.2$ & & \\
    1140856974 & 2025-09-08 17:28 & 2025-09-08 19:05 & 2373  & 2116 & $<7.53$ & $<662.2$  & & \\
    \hline
    \end{tabular}}
    \end{center}
    \tablenotetext{a}{The weighted-mean magnitudes and errors for each orbit, which have been corrected for Galactic extinction, specifically 0.378 mag in the $R$ band and 0.632 mag in the $B$ band. The data analysis uses a time bin of 5 minutes.}
    \tablenotetext{b}{The significance level of variability for each orbit observational data.}
    \tablenotetext{c}{The left numbers indicate the number of detection points with a time bin of 5 minutes, while the right numbers indicate the number of valid images for that orbit observation.}
    \tablenotetext{d}{``1--20'' corresponds to all the SVOM-VT observation data in Figure 1, while ``1--4'' and ``6--11'' correspond to the consecutive four and six orbit observations in Figure 2 respectively.}
    \tablenotetext{e}{The 3$\sigma$ upper limit of the MXT flux in the 0.3--10\,keV energy band.}
    \tablenotetext{f}{The 3$\sigma$ upper limit of the ECLAIRs flux in the 4--150\,keV energy band.}
\end{table*}

\clearpage
\appendix
\section{Multiwavelength Observations and Data Analysis}

\subsection{Swift-XRT}

Swift-XRT has conducted a total of 18 observations of PKS 1725+123. In this work, the X-ray spectral analysis of PKS 1725+123 is mainly based on 16 observations. The data from two observations (OBSIDs 00019638007 and 00019638009) were not utilized because the exposures were insufficient to acquire an adequate number of photons for spectral fitting. The data were processed using the XRT Data Analysis Software (XRTDAS, v.3.7.0) within the \texttt{HEASoft} package (v.6.34). The raw event files were calibrated using the calibration files from the Swift-XRT CALDB (v.20241028) through the \texttt{xrtpipeline} task. Spectra of source and background photons were extracted using the \texttt{xselect} task. We defined the source region as a circle with a radius of $47^{\prime\prime}$ and an annulus with inner and outer radii of $71^{\prime\prime}$ and $142^{\prime\prime}$, respectively, both centered on the coordinates of PKS 1725+123 provided in \citet{2019ApJS..242....5X}. Ancillary response files, incorporating cumulative exposure maps to account for point-spread function losses and CCD defects, were generated using the \texttt{xrtmkarf} task.

The spectra of PKS 1725+123 for different Swift-XRT observations were indicidually fitted with an absorbed PL model within \texttt{Xspec} \citep[v.12.14.1;][]{1999ascl.soft10005A}, specifically, \texttt{TBABS}$\times$\texttt{POWERLAW}. The PL model is described as follows: 
\begin{equation}
\frac{dN(E)}{dE}=N_{0}\left(\frac{E}{E_0}\right)^{-\Gamma_{\rm X}},
\end{equation}
where $N_{0}$ is the PL normalization, $E_{0}=1~{\rm keV}$ is the scale parameter of photon energy, and $\Gamma_{\rm X}$ is the photon spectral index. In this model, the \texttt{TBABS} component accounts for the Galactic photoelectric absorption. During the spectral fitting process, the column density ($N_{\rm H}$) was fixed at the Galactic value \citep[$N_{\rm H}=8.64\times10^{20}~{\rm cm^{-2}}$;][]{2016A&A...594A.116H} since the data quality of Swift - XRT is inadequate to constrain the intrinsic absorption. To improve the quality of the spectral fitting, the \texttt{grppha} task was utilized to re-bin the spectra, ensuring a minimum of 3 counts to guarantee the validity of the C-statistic during the spectral fitting. Moreover, a minimum of 20 counts was applied to use the $\chi^{2}$ statistic during the spectral fitting for the observation with sufficient photon counts. The best-fit parameters are summarized in Table \ref{tab:xrt}.

\subsection{Fermi-LAT}

In the Fermi-LAT 14-year Source Catalog (4FGL-DR4, \citealp{Ballet2024_4FGLDR4,2022ApJS..260...53A}), PKS 1725+123 is associated with $\gamma$-ray source 4FGL J1728.0+1216. We selected the Pass 8 data within the 0.1--1000 GeV band from a 15$^\circ$ region of interest (ROI) centered on the position of PKS 1725+123 (R.A.=262$^\circ$.029, decl.=12$^\circ$.261). The data cover the period from 2008 August 4 to 2025 October 1 (MJD 54682--60949). The binned maximum likelihood analysis was conducted using the publicly available software \textit{fermitools} (ver. 2.2.0, \citealp{2019ascl.soft05011F}) and $\rm P8R3\_SOURCE\_V3$ instrument response function (IRF), with a bin size of 0.1$^\circ$. Photons with zenith angles exceeding 90$^\circ$ were excluded to minimize the contamination of $\gamma$-rays emission from the Earth limb. All sources within the ROI were included in the model. The spectral parameters of the sources located within a circle with a radius of 6$^\circ$ were left free, while the other parameters were fixed at their 4FGL-DR4 values. The background emission was modeled using the diffuse Galactic interstellar emission (``$\rm gll\_iem\_v07.fits$'') and isotropic emission (``$\rm iso\_P8R3\_SOURCE\_V3\_V1.txt$''), and only their normalization parameters were kept free.

We use the maximum test statistic (TS) to evaluate the significance of the $\gamma$-ray signals of a source, where $\rm TS=2(log\mathcal{L}_{\rm src}-log\mathcal{L}_{\rm null})$, where $\mathcal{L}_{\rm src}$ and $\mathcal{L}_{\rm null}$ are the likelihood values for the background with and without a source. The threshold is set at TS=25 (corresponding to 5$\sigma$). The appearance of new sources in the ROI may affect the $\gamma$-ray spectrum and light curve of the research source. To eliminate the influence of new sources, we searched for possible new $\gamma$-ray sources within a 5$^\circ$ field of view and found no new sources beyond those listed in the 4FGL-DR4.

Through maximum likelihood fitting, the value of TS=2495.5 was obtained for 4FGL J1728.0+1216. The 17-year average $\gamma$-ray spectrum can be well explained in the LP form \citep{2004A&A...413..489M}, specifically, 
\begin{equation}
\frac{dN(E)}{dE}=N_0(\frac{E}{E_{\rm b}})^{-(\Gamma_{\gamma}+\beta\log(E/E_{\rm b}))},
\end{equation}
where $N(E)$ is the photon distribution as a function of energy, $\Gamma_{\gamma}$ is the photon spectral index, $E_{\rm b}$ is the scale parameter of photon energy, and $\beta$ is the curvature parameter. An average spectrum with $\Gamma_{\gamma}=2.16\pm0.03$ and $\beta=0.05\pm0.01$ was obtained, and the 17-year average flux is $F_{\rm 0.1-1000~GeV}=(2.12\pm0.10)\times10^{-11}$ erg cm$^{-2}$ s$^{-1}$. These results are roughly consistent with those of 4FGL-DR4. 

Using the same method, we obtained the time-resolved spectra for the three time intervals: 2025 August 16--18 (MJD 60903--60905), August 19--21 (MJD 60906--60908), and August 22--24 (MJD 60909--60911). Different from the 17-year integrated spectrum, the three time-resolved spectra can be well modeled by a simple PL function with a flatter photon spectral index. 

\begin{figure*}[ht!]
    \centering
    \includegraphics[width=1.1\textwidth]{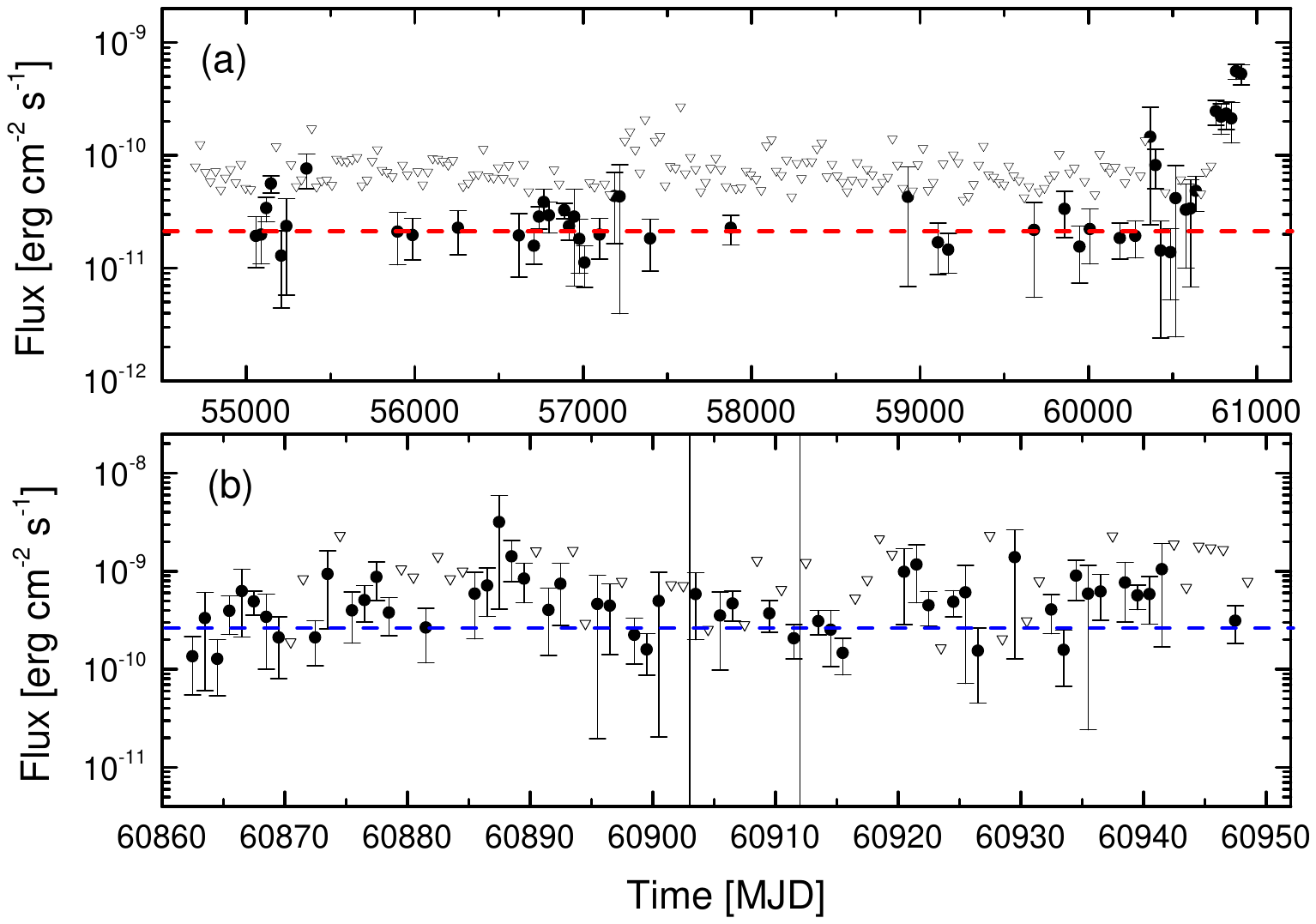}
     \caption{Light curves of PKS 1725+123 observed by the Fermi-LAT in the 0.1-1000 GeV band. If TS$<9$, an upper limit (represented by open inverted triangles) is provided for that time bin. Panel (a): the 17-year long-term light curve, derived with a time bin of 30 days. The red horizontal dashed line denotes the $\sim$17-year average flux, i.e., $F_{\rm 0.1-1000~GeV}=(2.12\pm0.10)\times10^{-11}$ erg cm$^{-2}$ s$^{-1}$. Panel (b): the 3-month light curve with a time bin of 1 day, covering the period from 2025 July 06 to 2025 October 01. The blue horizontal solid line denotes the weighted average flux of $2.65\times10^{-10}$erg cm$^{-2}$ s$^{-1}$, which is derived using Equation (1). The two black vertical lines represent the capture time interval of the three time-resolved spectra in Figure \ref{Sectrum_LAT}. }
      \label{lc-LAT}
\end{figure*}

We generated the 17-year long-term light curve using a time bin of 30 days, as displayed in Figure \ref{lc-LAT}(a). During the construction of the light curves, only the parameters of the target source and two background files were released, while the other parameters were fixed at the optimal fitting values. If TS$<9$, an upper limit of flux was provided for that time bin. We fixed the target source in the form of a PL function with a photon spectral index of 2 to calculate the upper limit value. Using the same method, we also generated the three-month (2025 July 06 to 2025 October 01) light curve using a time bin of 1 day, as depicted in Figure \ref{lc-LAT}(b). 

\section{SED Modeling}

\subsection{Model}

As described in Section \ref{resultsec:svom-vt}, SVOM-VT has observed optical variability of PKS 1725+123 with a timescale of less than an hour, which suggests a small size of the emission region. Meanwhile, the hard optical spectrum, in combination with a soft X-ray spectrum, indicates a very high synchrotron peak frequency. Further considering a very hard Fermi-LAT spectrum and a high redshift, a typical one-zone leptonic model fails to provide an adequate explanation for the broadband SED of PKS 1725+123. Nevertheless, during the optical flare state observed by SVOM-VT, the emission in the X-ray and $\gamma$-ray bands also demonstrates a high-flux state, indicating that the emission among the three bands has some kind of connection. In other words, neither the model with two independent zones nor the hadronic model is suitable (e.g., \citealt{2026arXiv260501601C}). Additionally, the detection of VHE emission require extremely high-energy electrons to produce it. It is noted that both radio observations and numerical simulations have indicated that the jets possess a spine-sheath velocity stratification structure (e.g., \citealt{2008A&A...488..795R, 2017MNRAS.466.3544C, 2018ApJ...855..128W}). When electrons pass through the shear layers associated with velocity gradients, the shear acceleration process, a Fermi-type mechanism similar to shock or stochastic acceleration, can produce a population of higher-energy electrons \citep{2004ApJ...617..155R, 2007Ap&SS.309..119R, 2017ApJ...842...39L, 2019Galax...7...78R}. Therefore, a two-zone leptonic model, as described in Section \ref{sec:SED}, is proposed.

The electron distribution in both emission regions is taken as a PL, characterized by an electron number density parameter $N_0$, a spectral index $p$, and a minimum ($\gamma_{\min}$) and a maximum ($\gamma_{\max}$) Lorentz factor of electrons. Both radiation regions are assumed to be spherical with a radius $R$, a magnetic field strength $B$, and a Doppler boosting factor $\delta$, where $\delta=1/(\Gamma-\sqrt{\Gamma^2-1}\cos\theta)$, and $\Gamma$ and $\theta$ are the bulk Lorenz factor and viewing angle of the emission region. The value of $R$ is constrained by the variability timescale, specifically, $R=\Delta t c\delta/(1 + z)$. As displayed in Figure \ref{lc-VT_hour}, a continuous decrease in flux was observed during the fourth-orbit observation (a similar phenomenon was also presented in the VT$\_B$ band of the second-orbit observation), with a duration of 30 minutes from the first to the last data points. Therefore, a $\Delta t$ value of 30 minutes is adopted for the compact region, and it is assumed that $\delta=\Gamma$. For the extended region, a $\Delta t$ value of 1 day is adopted, and the same viewing angle as that of the compact region is assumed. Given that the VHE $\gamma$-rays are detected during this high-flux state, it is assumed that the emission region is not located deep within the broad-line regions (BLRs), as this would result in strong attenuation through $\gamma\gamma$-annihilation (e.g., \citealt{2006ApJ...653.1089L, 2010ApJ...717L.118P}). Therefore, the energy density of the BLR is fixed at 0.5 times the typical value of $2.73 \times 10^{-2}\ \Gamma^2\ \mathrm{erg\ cm^{-3}}$, and the spectrum of the BLR can be approximated by a blackbody with a peak at $2 \times 10^{15}\ \Gamma\ \mathrm{Hz}$ in the co-moving frame (e.g., \citealt{2008MNRAS.386..945T, 2014ApJS..215....5K}). The electron radiative processes have incorporated synchrotron self-absorption \citep{1979rpa..book.....R}, the Klein–Nishina effect \citep{1998ApJ...509..608T}, and extragalactic background light absorption \citep{2022ApJ...941...33F}.

Since it is assumed that the two regions overlap, the code also incorporates the emission resulting from the interaction of the two zones. The synchrotron radiation from the two regions provide an additional target photon field for each other for the IC scattering. Moreover, due to the relative motion between the compact region and the extended region, the energy density of the seed photons will be amplified \citep{2005A&A...432..401G, 2024ApJS..271...10W}. However, the radiation contribution from the interaction of the two zones is mainly attributed to the synchrotron photons from the compact region undergoing IC scattering by the relativistic electrons within the extended region. In contrast, the flux contribution from the opposite process can be disregarded.

The fitting results are presented in Figure \ref{SED} and the fitting parameters are given in Table \ref{Tab: Parameters}. It should be noted that the derived parameter values are based on visual assessments and the model parameters cannot be fully constrained by the current observational data.

\subsection{Cooling and Acceleration Timescales of Electrons}

Given the rapid variability observed in the optical band, it was assumed that there is a compact emission region during the SED modeling. The synchrotron frequency ($\nu_{\rm s}$) averaged over the spectral shape for an electron with Lorentz factor ($\gamma_{\rm e}$) is $\nu_{\rm s}=3.7\times10^6\gamma_{\rm e}^2B\frac{\delta}{1+z}$ \citep{1979rpa..book.....R}. The cooling timescale of electrons through the synchrotron process is estimated by $t_{\rm cool}=\frac{6\pi m_{\rm e} c}{\sigma_{\rm T} \gamma_{\rm e}B^2}$ \citep{1979rpa..book.....R,1999MNRAS.306..551C}, where $m_{\rm e}$ is the electron mass and $\sigma_{\rm T}$ is Thomson cross-section. For the contribution to the optical emission, the cooling timescale of electrons is about 46 minutes in the co-moving frame and several minutes in the observer frame. 

We also estimate the shear-acceleration timescale of electrons in the extended zone under the strong-scattering regime. The acceleration timescale is given by the following equations \citep{2017ApJ...842...39L}:
\begin{equation}
t_{\mathrm{acc}}=\frac{15}{6-q}A^{-2}\tau^{-1},
\end{equation}
\begin{equation}
A = \Gamma_{\rm jet}^2 (\partial \beta_{\rm jet}/\partial r)c,
\end{equation}
\begin{equation}
\tau = \lambda/c,
\end{equation}
\begin{equation}
\lambda = \xi^{-1} r_g (r_g/\Lambda_{\max})^{1-q},
\end{equation}
where $\Gamma_{\rm jet}=\Gamma=35$ denotes the bulk Lorentz factor of the jet, $\beta_{\rm jet}$ represents the speed of the jet in units of $c$, and $r_{\rm g}$ is the gyroradius of electrons. In this scenario, we adopt $\xi \sim 0.1$, which represents the ratio of turbulent to ordered magnetic-field energy density. We employ the Kolmogorov turbulence model ($q=5/3$; \citealt{1941DoSSR..30..301K}) and assume an acceleration region width $\Delta r \sim 10^{15}$ cm with $\Lambda_{\max} \sim \Delta r$. Then, we obtain $t_{\mathrm{acc}} \sim 4\times10^3$ s.

It should be noted that the estimation results for the cooling and acceleration timescales of electrons are consistent with the observations.

\subsection{Jet Power}

By assuming that the the jet power is carried by relativistic electrons ($P_{\rm e}$), magnetic fields ($P_{B}$), and radiation ($P_{\rm r}$), i.e., $P_{\rm jet} = \pi R^2 \Gamma^2 c (U_{\rm e} + U_B + U_{\rm r})$, and using the SED fitting parameters, we calculate the jet power and the powers of each component for both the compact and extended regions. $U_e$, $U_B$, and $U_r$ are the energy densities of relativistic electrons, magnetic fields, and radiation in the co-moving frame, respectively, which are given by the following equations \citep[e.g.,][]{2008MNRAS.385..283C, 2010MNRAS.402..497G}:
\begin{equation}
U_{\rm e} = m_{\rm e} c^2 \int N(\gamma)\gamma d\gamma ,
\end{equation}
\begin{equation}
U_B = \frac{B^2}{8\pi},
\end{equation}
\begin{equation}
U_{\rm r} = \frac{L_{\rm bol}}{4\pi R^2 c \delta^4},
\end{equation}
where $L_{\rm bol}$ is the bolometric luminosity.

\begin{figure*}[ht!]
    \centering
    \includegraphics[width=1\textwidth]{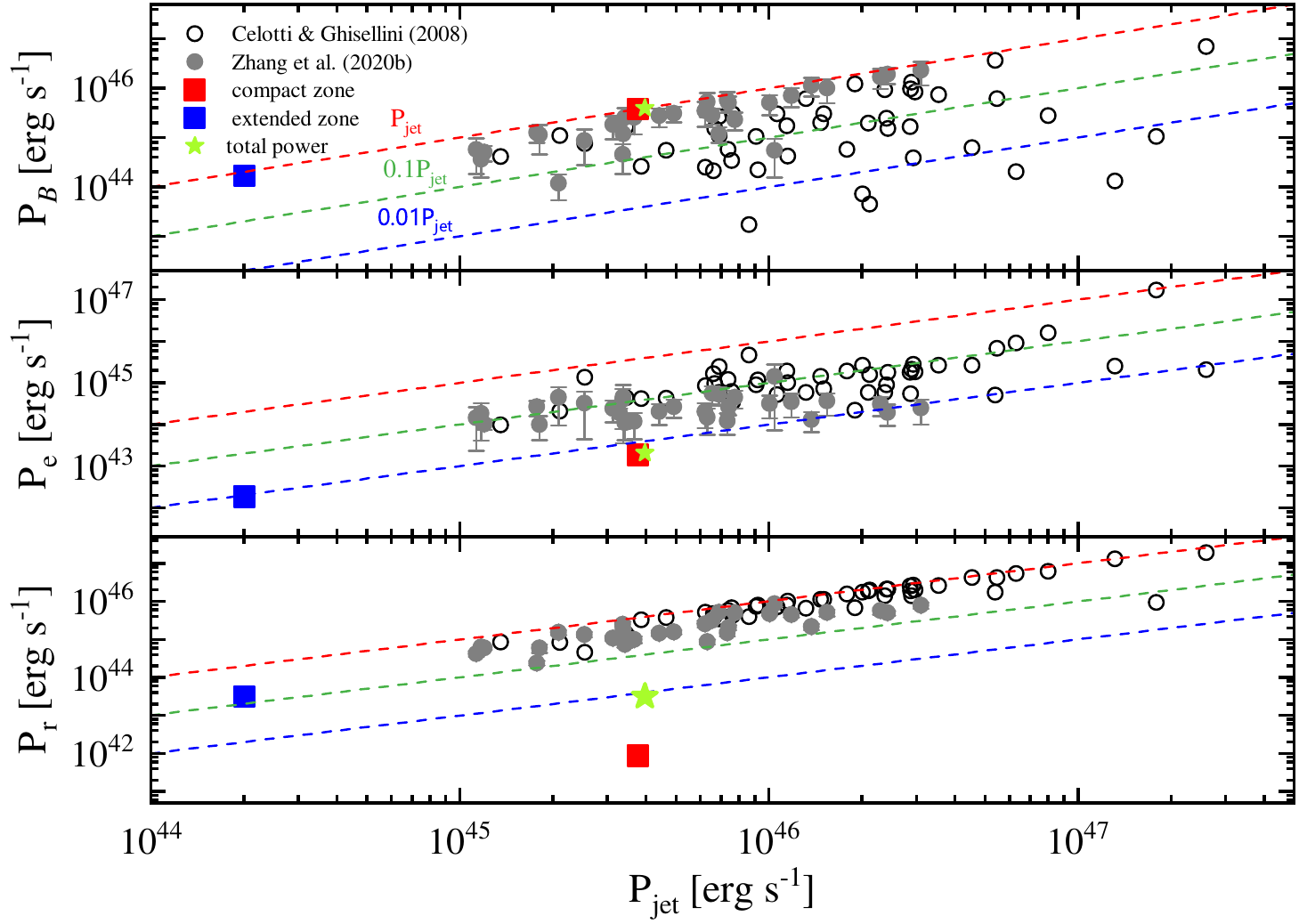}
     \caption{$P_B$, $P_{\rm e}$, and $P_{\rm r}$ as functions of $P_{\rm jet}$. The red and blue solid squares denote the values for the compact and extended zones of PKS 1725+123, respectively; meanwhile, the green solid star represents the power sum of the two zones. The black open circles and gray solid circles represent the data of a $\gamma$-ray emission FSRQ sample from \cite{2008MNRAS.385..283C} and \cite{2020ApJ...899....2Z}, respectively.}
      \label{pjet}
\end{figure*}

The derived jet power and the powers of each component for the compact and extended regions are presented in Table~\ref{Tab: Parameters}. It is found that for the compact zone, $P_B/P_{\rm jet} \sim 0.995$ and $P_{\rm r}/P_{\rm jet} \sim 0.0002$, and for the extended zone, $P_B/P_{\rm jet} \sim 0.841$ and $P_{\rm r}/P_{\rm jet} \sim 0.150$. These results suggest that both zones are highly magnetized; however, the compact zone exhibits low radiation efficiency while the extended zone demonstrates high radiation efficiency. We compare the derived jet powers of PKS 1725+123 with a $\gamma$-ray emission FSRQ sample from \cite{2008MNRAS.385..283C} and \cite{2020ApJ...899....2Z}, and we plot $P_{\rm e}$, $P_B$, and $P_{\rm r}$ as functions of $P_{\rm jet}$ in Figure~\ref{pjet}. We combine the compact (red solid square) and extended (blue solid square) zones as a single jet emission region of PKS 1725+123, which is marked as a green solid star in Figure~\ref{pjet}. Comparing with other $\gamma$-ray emission FSRQs, on average, PKS 1725+123 exhibits a low jet power and radiation efficiency, but a highly magnetized jet.

\clearpage

\begin{table*}
    \begin{center}
    \caption{Swift-XRT Spectral Analysis Results for PKS 1725+123}
    \label{tab:xrt}
    \resizebox{\textwidth}{!}{
    \begin{tabular}{ccccccccc}
    \hline
    \hline
    OBSID & Date & Exposure & $N_{0}$ & $\Gamma_{\rm X}$ & Fit Statistic & $F_{0.3-10}$ & $F_{0.3-2}$ & $F_{2-10}$ \\
    \cmidrule(r){7-9}
    & (YYYY-MM-DD) & (s) & ($10^{-4}~{\rm ph~cm^{-2}~s^{-1}~keV^{-1}}$) & & & \multicolumn{3}{c}{($10^{-12}~{\rm erg~cm^{-2}~s^{-1}}$)} \\
    \hline
    00019638001 & 2025-03-21 & 1296 & $3.28^{+0.71}_{-0.63}$ & $1.59\pm0.25$ & 10/13 & $2.52^{+0.37}_{-0.40}$ & $0.92\pm0.14$ & $1.60^{+0.24}_{-0.25}$ \\
    00019638002 & 2025-03-24 & 1486 & $3.10^{+0.85}_{-0.72}$ & $2.09^{+0.40}_{-0.38}$ & 10/7 & $1.66^{+0.32}_{-0.35}$ & $0.97^{+1.09}_{-0.20}$ & $0.70^{+0.13}_{-0.14}$ \\
    00019638004 & 2025-03-29 & 438 & $2.46^{+2.34}_{-1.32}$ & $1.42^{+1.01}_{-0.99}$ & 4/1 & $2.27^{+0.72}_{-0.80}$ & $0.67^{+0.21}_{-0.24}$ & $1.59^{+0.50}_{-0.56}$ \\
    00019638005 & 2025-08-19 & 1509 & $16.03^{+1.16}_{-1.18}$ & $2.22\pm0.13$ & ${5/9}^{\dag}$ & $8.17^{+0.54}_{-0.58}$ & $5.19^{+0.35}_{-0.37}$ & $2.98^{+0.20}_{-0.21}$ \\
    00019638006 & 2025-08-21 & 837 & $17.55^{+1.73}_{-1.74}$ & $2.49^{+0.22}_{-0.21}$ & ${4/4}^{\dag}$ & $8.49^{+0.76}_{-0.83}$ & $6.28^{+0.56}_{-0.62}$ & $2.22^{+0.20}_{-0.22}$ \\
    00019638010 & 2025-08-25 & 1294 & $5.24^{+0.85}_{-0.78}$ & $1.76\pm0.21$ & 20/22 & $3.46^{+0.42}_{-0.44}$ & $1.51^{+0.18}_{-0.19}$ & $1.95^{+0.24}_{-0.25}$ \\
    00019638011 & 2025-08-26 & 2202 & $7.40^{+1.01}_{-0.99}$ & $1.95^{+0.22}_{-0.21}$ & ${3/4}^{\dag}$ &$4.28^{+0.38}_{-0.42}$ & $2.22^{+0.20}_{-0.22}$ & $2.05^{+0.18}_{-0.20}$ \\
    00019638012 & 2025-09-07 & 411 & $8.00^{+1.82}_{-1.61}$ & $1.41^{+0.28}_{-0.29}$ & 8/11 & $7.38^{+1.16}_{-0.23}$ & $2.20^{+0.35}_{-0.37}$ & $5.18^{+0.82}_{-0.86}$ \\
    00019638013 & 2025-09-10 & 870 & $5.52^{+1.07}_{-0.95}$ & $1.52\pm0.24$ & 15/16 & $4.52^{+0.60}_{-0.63}$ & $1.54^{+0.20}_{-0.21}$ & $2.98^{+0.40}_{-0.42}$ \\
    00019638014 & 2025-09-11 & 865 & $4.45^{+0.99}_{-0.86}$ & $1.09\pm0.23$ & 14/16 & $6.07^{+0.80}_{-0.84}$ & $1.21^{+0.16}_{-0.17}$ & $4.86^{+0.64}_{-0.67}$ \\
    00019638015 & 2025-09-12 & 992 & $2.62^{+1.16}_{-0.93}$ & $1.15^{+0.51}_{-0.53}$ & 4/4 & $3.29^{+0.84}_{-0.94}$ & $0.72^{+0.18}_{-0.20}$ & $2.57^{+0.65}_{-0.73}$ \\
    00019638016 & 2025-09-13 & 1075 & $6.45^{+1.01}_{-0.92}$ & $1.38\pm0.18$ & 38/27 & $6.13^{+0.65}_{-0.68}$ & $1.77^{+0.19}_{-0.20}$ & $4.36^{+0.46}_{-0.48}$ \\
    00019638017 & 2025-09-14 & 1075 & $6.40^{+0.97}_{-0.88}$ & $1.69^{+0.20}_{-0.19}$ & 19/23 & $4.46^{+0.51}_{-0.53}$ & $1.81\pm0.21$ & $2.65^{+0.30}_{-0.31}$ \\
    00019638018 & 2025-09-28 & 2316 & $3.91^{+0.51}_{-0.48}$ & $1.59\pm0.16$ & 36/32 & $3.01^{+0.30}_{-0.31}$ & $1.10\pm0.11$ & $1.91\pm0.19$ \\
    00019638019 & 2025-09-29 & 1810 & $3.34^{+0.55}_{-0.50}$ & $1.59\pm0.20$ & 36/21 & $2.57^{+0.31}_{-0.32}$ & $0.94^{+0.11}_{-0.12}$ & $1.63^{+0.20}_{-0.21}$ \\
    00019638020 & 2025-10-01 & 1725 & $3.31^{+0.57}_{-0.52}$ & $1.32\pm0.19$ & 18/24 & $3.38^{+0.39}_{-0.40}$ & $0.91^{+0.10}_{-0.11}$ & $2.47^{+0.28}_{-0.30}$ \\
    \hline
    \end{tabular}}
    \end{center}
    \tablenotetext{$\dag$}{The $\chi^{2}$ statistic was employed during the spectral fits of these observations.}
\end{table*}

\begin{deluxetable*}{lccr} 
    \renewcommand{\arraystretch}{1.1} 
    \setlength{\tabcolsep}{10pt}  
    \label{Tab: Parameters}
    \tablecaption{Parameters of SED Fitting and Jet Powers}
    \tablehead{\colhead{Parameter}&\colhead{Symbol}& \colhead{Compact Zone}& \colhead{Extended Zone}}
    \startdata        
    Electron Density Parameter &$N_0$ [cm$^{-3}$] & $6\times10^{2}$ & $1\times10^{2}$ \\
    Electron Spectral Index  &$p$  &$1.4$   &   $2$ \\
    Minimum Lorentz Factor for Electrons&$\gamma_{\rm min}$ & $1\times10^{2}$  &$3\times10^{3}$ \\
    Maximum Lorentz Factor for Electrons&$\gamma_{\rm max}$   & $4.5\times10^{3}$  & $5\times10^{5}$\\
    Radius of Emission Region&  $R$ [cm]   & $1.2\times10^{15}$    &$1.9 \times 10^{16}$\\
    Bulk Lorentz Factor      &$\Gamma$   & $35$   &$5.9$\\
    Doppler Factor&$\delta$   & $35$   &$11.4$\\
    Magnetic Field&$B$ [G]   & $24 $ & $1 $\\
    Energy Density of BLR&$U_{\rm BLR}$ [erg cm$^{-3}$]    & $1.36\times10^{-2}$   &$1.36\times10^{-2}$\\
    Equipartition Ratio &$U_B/U_e$     & $ 2 \times 10^{2}$    &$9.5 \times 10^{1}$ \\ 
    Radiative Power               &  $P_{\rm r}$ [erg s$^{-1}$]    & $ 8.58 \times 10^{41}$    &$3.02 \times 10^{43}$ \\ 
    Nonthermal Electron Power        &  $P_{\rm e}$ [erg s$^{-1}$]    & $ 1.87 \times 10^{43}$    &$1.77 \times 10^{42}$ \\ 
    Magnetic Field Power                &  $P_B$ [erg s$^{-1}$]          & $ 3.75 \times 10^{45}$    &$1.69 \times 10^{44}$ \\    
    Jet Power               &  $P_{\rm jet}$ [erg s$^{-1}$]  & $ 3.77 \times 10^{45}$    &$2.01 \times 10^{44}$ \\     
    \noalign{\smallskip}
    \enddata
\end{deluxetable*}

\end{document}